\documentclass[sigconf]{acmart}
\AtBeginDocument{%
  \providecommand\BibTeX{{%
    \normalfont B\kern-0.5em{\scshape i\kern-0.25em b}\kern-0.8em\TeX}}}

\copyrightyear{2024}
\acmYear{2024}
\setcopyright{rightsretained}
\acmConference[UIST '24]{The 37th Annual ACM Symposium on User Interface
Software and Technology}{October 13--16, 2024}{Pittsburgh, PA, USA}
\acmBooktitle{The 37th Annual ACM Symposium on User Interface Software and
Technology (UIST '24), October 13--16, 2024, Pittsburgh, PA, USA}
\acmDOI{10.1145/3654777.3676410}
\acmISBN{979-8-4007-0628-8/24/10}
\acmSubmissionID{6869}

\usepackage{booktabs}
\usepackage{multirow}
\usepackage{cleveref}

\usepackage{enumitem}
\setlist{noitemsep,parsep=0pt,partopsep=0pt, leftmargin=10pt} 
\usepackage{listings}

\begin{document}

\title{NotePlayer: Engaging Jupyter Notebooks for Dynamic Presentation of Analytical Processes}

\author{Yang Ouyang}
\email{ouyy@shanghaitech.edu.cn}
\orcid{0009-0000-5841-7659}
\affiliation{%
  \institution{ShanghaiTech University}
  \country{Shanghai, China}
}
\author{Leixian Shen}
\email{lshenaj@connect.ust.hk}
\orcid{0000-0003-1084-4912}
\affiliation{%
  \institution{The Hong Kong University of Science and Technology}
  \country{Hong Kong SAR, China}
}
\author{Yun Wang}
\email{wangyun@microsoft.com}
\orcid{0000-0003-0468-4043}
\affiliation{%
  \institution{Microsoft}
  \country{Beijing, China}
}
\authornotemark[1]
\author{Quan Li}
\email{liquan@shanghaitech.edu.cn}
\orcid{0000-0003-2249-0728}
\affiliation{%
  \institution{ShanghaiTech University}
  \country{Shanghai, China}
}
\authornotemark[0]
\authornote{Quan Li and Yun Wang are the corresponding authors.}

\begin{abstract}

Diverse presentation formats play a pivotal role in effectively conveying code and analytical processes during data analysis. One increasingly popular format is tutorial videos, particularly those based on Jupyter notebooks, which offer an intuitive interpretation of code and vivid explanations of analytical procedures. 
However, creating such videos requires a diverse skill set and significant manual effort, posing a barrier for many analysts.
To bridge this gap, we introduce an innovative tool called \textit{NotePlayer}, which connects notebook cells to video segments and incorporates a computational engine with language models to streamline video creation and editing. Our aim is to make the process more accessible and efficient for analysts. To inform the design of \textit{NotePlayer}, we conducted a formative study and performed content analysis on a corpus of 38 Jupyter tutorial videos. This helped us identify key patterns and challenges encountered in existing tutorial videos, guiding the development of \textit{NotePlayer}. Through a combination of a usage scenario and a user study, we validated the effectiveness of \textit{NotePlayer}. The results show that the tool streamlines the video creation and facilitates the communication process for data analysts.
\end{abstract}



\begin{teaserfigure}
\end{teaserfigure}

\newcommand{\writing}{\textcolor{red}}
\newcommand{\ouyang}{\textcolor{black}}
\newcommand{\revisionyang}{\textcolor{black}}

\maketitle
\section{Introduction}

\par Effective communication in data analysis is essential for translating complex codes and analytical processes into understandable formats~\cite{li2023notable,chevalier2018analysis}. Employing diverse presentation formats, ranging from traditional text reports to interactive digital platforms, is crucial in achieving this objective~\cite{kery2018story,lin2023inksight,wang2023slide4n}. Among these methods, tutorial videos have emerged as particularly popular, offering an intuitive interpretation of codes while vividly articulating underlying analytical processes~\cite{weeks2017evaluating,Bowles-Terry2010Best,bao2018vt}.
\par In parallel, the Jupyter Notebook platform has gained significant recognition within the programming and data analysis community~\cite{rule2018aiding,ramasamy2023visualising,li2023notable,wang2024outlinespark}. Regarded as a vital tool for efficient data analysis, its structured design, featuring input and output cells, facilitates iterative exploration and provides instant feedback, essential for hypothesis testing and comprehension~\cite{chattopadhyay2020s,rule2018exploration,head2019managing,li2023notable}. Notably, the user interface of Jupyter Notebook frequently appears in a variety of video tutorials, making it accessible on online platforms such as YouTube, MOOCs, and Khan Academy~\cite{barba2019teaching,wagemann2022five,snickars2009youtube,saadatdoost2016understanding,thompson2011khan}. \revisionyang{Many findings highlight the benefits of tutorial videos in enriching both theoretical understanding and practical application in data science~\cite{kross2019practitioners,guo2014video,lu2017exploring,wang2019creating}.}

\par 
However, creating tutorial videos based on Jupyter notebooks poses challenges for data analysts. Effective programming tutorial videos demand a blend of skills, including providing insightful guidance and understanding viewer engagement~\cite{Bowles-Terry2010Best,macleod2017documenting,weeks2017evaluating}. Balancing the clarity and structure of the original analysis with engaging and informative content is crucial~\cite{raj2018role,kim2017pedagogical}. Moreover, many analysts may lack familiarity with specialized video editing tools like Adobe After Effects and Premier. While powerful, these tools are not tailored for programming tutorial videos, adding complexity, especially for those eager to share their analytical processes but inexperienced in video editing. Despite efforts to enhance presentation and interactivity within Jupyter notebooks~\cite{kang2021toonnote,wang2023slide4n,li2023notable,lin2023inksight}, a gap remains in fully capturing the analytical process. These efforts often prioritize showcasing insights over providing a detailed walkthrough of the analytical journey. Consequently, viewers may lack a clear understanding of the methodologies and critical thinking behind the findings.

\par 
To address these challenges, we initially conducted an interview-based formative study involving four creators to gain insights into their workflows and identify challenges in tutorial video creation. Subsequently, we conducted a content analysis of 38 programming tutorial videos to explore their characteristics and detect common patterns, thus shedding light on the current landscape of this media format. The formative study revealed several challenges in creating effective videos, including maintaining a clear logical flow aligned with the code structure, enhancing comprehension through visual aids, and the time-consuming nature of editing to rectify errors and enhance clarity. The content analysis highlighted that these videos employ various narration styles (informative, structural, interactive) along with specific creator behaviors (e.g., annotating). Key strategies for improving tutorials include emphasizing code snippets through annotations, ensuring smooth content flow, and engaging viewers with interactive questions. We have also arrived at a consensus on transitioning from static screen recordings to more interactive and engaging methods, thereby providing comprehensive insights into analytical workflows. These insights have informed the development of a set of design considerations to steer the future development of our tool.
\par 
\par Based on our findings, we present \textit{NotePlayer}, an authoring tool seamlessly integrating notebook cells with video segments. Powered by a computational engine, \textit{NotePlayer} simplifies the creation and editing of videos. Initially, \textit{NotePlayer} extracts data, encompassing both input and output cells, from the user's original notebook. This process effectively demonstrates the logical flow inherent in the analytical process by analyzing code cells. Users can then navigate through this logical flow, confirming each scene as they progress. Within each scene, the tool generates initial narration by leveraging detailed code cell information and essential elements, such as specific code snippets and corresponding user annotations. This process utilizes the robust natural language understanding capabilities of Large Language Models. Additionally, after fine-tuning the narration and integrating it with specific visual elements and preset settings, such as animations and layouts, an initial dynamic presentation is crafted. Supporting this process is the finalization of a "design script" that organizes the streaming content into coherent scenes, seamlessly integrating code, annotations, and visuals. To ensure accessibility, text-to-speech technologies are utilized to automatically generate audio narrations. Furthermore, \textit{NotePlayer} facilitates iterative refinement of streaming content, enabling users to edit their presentations, adjust annotations, and enhance narrations. This iterative process enhances the quality and consistency of the streaming media.

\par In evaluating \textit{NotePlayer}, we conducted a user study and received positive feedback regarding its usability and learnability. Participants highlighted that \textit{NotePlayer} significantly streamlines the creation of expressive Jupyter notebook tutorial videos, providing clear guidance and ease of learning, thus improving their communication process. \revisionyang{We view the integration of tutorial videos as a valuable supplemental tool that complements traditional learning methods. Our method provides additional support, effectively demystifying complex topics, increasing engagement, and offering flexible learning options.} 
The contributions of this work are summarized as follows:
\begin{itemize}
\item We conduct a formative study and content analysis to gain a comprehensive understanding of the general workflow, challenges, and common patterns involved in existing programming tutorial video creation practices.
\item We create \textit{NotePlayer}, a tool seamlessly connecting notebook cells with video segments, employing a computational engine embedded with language models to facilitate flexible video creation and editing.
\item We showcase the effectiveness of \textit{NotePlayer} through a user study, illuminating the tool's advantages and limitations.
\end{itemize}

\section{Related Work}
\par We examine previous works from two angles: Programming Tutorial Videos and Storytelling with Computational Notebooks.

\subsection{Programming Tutorial Videos}
\par Tutorials are extensively utilized as a platform for sharing coding and programming expertise among programmers~\cite{head2020composing}. They commonly include elements such as source code, textual explanations, code examples, and multimedia components, encompassing images and videos~\cite{tiarks2014does,mysore2017torta}.
\par Programming tutorial videos play a vital role in providing an intuitive understanding of code and explaining analytical processes clearly. These videos are commonly found on formal online platforms such as YouTube~\cite{snickars2009youtube}, MOOCs~\cite{saadatdoost2016understanding}, and Khan Academy~\cite{thompson2011khan}. 
\revisionyang{Recent studies support the effectiveness of tutorial videos in data science education. Kross et al.~\cite{kross2019practitioners} discussed the challenges faced by data science instructors in integrating code, data, and communication within teaching workflows, emphasizing the importance of effective communication. Guo et al.~\cite{guo2014video} showed that interactive and concise videos significantly boost student engagement and understanding, essential for clarifying complex data science techniques and coding practices. Moreover, Lu et al.~\cite{lu2017exploring} investigated the impact of tutorial videos on teaching practical skills, while Wang et al.~\cite{wang2019creating} confirmed their effectiveness in increasing engagement in Computer Science and Engineering courses. These findings highlight the benefits of tutorial videos in enriching both theoretical understanding and practical application in data science.}  
\par Creating effective programming tutorial videos requires a combination of technical skills, educational insights, and an understanding of viewer engagement~\cite{macleod2017documenting,kim2017pedagogical,weeks2017evaluating}. Bowles-Terry et al.~\cite{Bowles-Terry2010Best} conducted research on learner preferences for online video tutorials, establishing a foundational understanding of learners' viewpoints in tutorial creation. Subsequently, Weeks et al.~\cite{weeks2017evaluating} furthered the discourse on video tutorial development by exploring optimal practices, emphasizing the critical role of usability, findability, and pedagogical efficiency in tutorial design.

\par Numerous studies have delved into the development of innovative tools for crafting tutorial videos. For instance, Mysore et al.~\cite{mysore2017torta} introduced the Torta system, which automates the creation of mixed-media tutorials for both GUI and command-line applications by tracing activities across the operating system. Bao et al.~\cite{bao2018vt} presented VT-Revolution, an interactive system aimed at enhancing the creation and viewing experience of programming video tutorials. Additionally, various studies have aimed to enhance interactivity, context, and clarity within programming tutorials. Khandwala et al.~\cite{khandwala2018codemotion} developed Codemotion, a tool focused on enriching interactivity in tutorial videos, highlighting the potential of interactive elements in making tutorials more engaging and effective for learners. Similarly, Buffardi et al.~\cite{Buffardi2022Integrating} explored integrating videos with programming practice, underscoring the importance of contextual and relatable content through the inclusion of diverse perspectives in tutorial videos. Moreover, Yadid et al.~\cite{Yadid2016Extracting} introduced a novel approach for extracting code from programming tutorial videos, emphasizing the significance of clarity in code presentation for enhanced learner comprehension.

\par Our focus is on enhancing communication in the realm of data analysis within the Jupyter Notebook environment, which has often been overlooked in traditional programming tutorial videos~\cite{schwichow2016students}. Specifically, there is a gap in understanding the unique workflows of data analysts as they create programming tutorial videos within Jupyter notebooks. To bridge this gap, our work delves into the challenges faced during the creation of such videos and seeks opportunities to streamline this process.

\subsection{Storytelling with Computational Notebooks}
\par This section delves into the Computational Notebook, a widely adopted platform extensively utilized by data analysts for their day-to-day tasks~\cite{jhome}. Each notebook is organized with numerous input and output cells that facilitate code editing and the display of results~\cite{kluyver2016jupyter}. The integrated interface for code and results aligns well with the needs of data analysis, supporting iterative coding and immediate result review~\cite{head2019managing,kery2018story}. Despite the benefits that data analysts derive from computational notebooks, these tools also present limitations. Rule et al.~\cite{rule2018exploration} highlighted the conflict between exploration and structured explanation within notebooks. Chattopadhyay et al.~\cite{chattopadhyay2020s} identified nine significant concerns, including analysis, code management, and collaboration issues. These limitations can hinder effective storytelling within computational notebooks.

\par Recent research has made significant advancements in improving storytelling within computational notebooks, introducing innovative tools and methods for enhancing data analysis and presentation~\cite{mooers2021modernizing,chattopadhyay2023make,harden2022exploring,o2015computational,kery2018interactions,PyGWalker}. We categorize these enhancements into two main sections: \textit{Analysis Enhancements} and \textit{Communication Enhancements}. In terms of \textit{Analysis Enhancements}, our focus is on tools that enhance user engagement and facilitate collaboration~\cite{weinman2021fork,wang2020callisto,wang2020assessing}. For instance, Head et al.\cite{head2019managing} introduced methods for gathering code within notebooks to manage messy code structures. Wang et al.\cite{wang2019data} presented a Jupyter extension designed to foster discussions around notebooks and facilitate collaborations. When it comes to \textit{Communication Enhancements}, the emphasis is on novel approaches to presenting and conveying analytical processes~\cite{wang2022documentation,wang2022stickyland}. Wenskovitch et al.\cite{wenskovitch2019albireo} introduced Albireo, a tool that visually summarizes notebook structures, aiding in the exploration of complex data stories. Additionally, NB2Slides\cite{zheng2022telling} and Slide4N~\cite{wang2023slide4n} automated the extraction of key points from notebook cells, organizing them into presentation slides for clearer communication. Furthermore, Li et al.\cite{li2023notable} introduced Notable, an on-the-fly assistant that enhances data documentation and organization, improving the clarity of data narratives. Lin et al.\cite{lin2023inksight} presented InkSight, a plugin that enhances chart documentation by allowing users to intuitively sketch their insights directly on visualizations. Together, these efforts enrich the experience of data analysts using notebooks, facilitating the creation of more engaging and collaborative storytelling within notebooks.
\par However, existing efforts have predominantly concentrated on reporting data facts or findings, neglecting the potential of streaming as a pivotal presentation medium capable of integrating narration and animation effects. In response, our work is oriented towards embracing conventional practices seen in programming tutorial videos. We take strides towards enabling a fluid and dynamic portrayal of the entire analysis process within computational notebooks, thereby enhancing communication.

\section{FORMATIVE STUDY}
Our study begins by examining the prevalent practices among data analysts in communicating their analytical processes, primarily through jupyter notebooks, to their audience.

\subsection{Participants and Procedure}
\label{sec:bottlenecks}
\par We conducted semi-structured interviews with four individuals, identified as \textbf{P1-P4}, all of whom are practitioners in Data Science (DS). They all have created tutorial notebook videos during the epidemic. Each participant brings over a decade of experience in computational data processing, along with several years of expertise specifically with Jupyter notebooks. The interviews with each participant were comprehensive, lasting approximately $30$ minutes.

\subsection{Data Analysis}
\par We conducted thematic analysis~\cite{clarke2017thematic} on the interview data and constructed an affinity diagram to explore the patterns of data analysts' workflows and the themes of challenges they encountered. Two researchers independently analyzed and open-coded the transcribed interview responses. Any discrepancies in the coding process were addressed through discussion and reconciliation to ensure consistency and accuracy in representing the participants' perspectives. Subsequently, the researchers utilized affinity diagramming to categorize the initial codes onto cards. Through iterative discussion and organization of the codes, several recurring patterns and themes were identified from the collected data.

\subsection{Key Findings}
\subsubsection{General Workflow}
\par According to the insights obtained from the interviews, the workflow of creating Jupyter Notebook tutorial videos involves five main steps:

\begin{itemize}
    \item \textbf{[S1]} \textbf{Planning}: In this initial stage, the data analyst carefully plans the content and structure of the notebook. This often includes removing irrelevant code cells and adding explanatory comments for clarity.
    \item \textbf{[S2]} \textbf{Recording the tutorial}: After planning, the data analyst proceeds to capture the live coding session using screen recording software within the computational notebook environment. This step provides the core visual aspect for the audience, demonstrating the code-writing process in detail, step-by-step.
    \item \textbf{[S3]} \textbf{Providing oral narration}: As the code unfolds on the screen, the data analyst simultaneously provides an oral narrative that complements the visual content. This narration goes beyond mere commentary, offering a detailed explanation of the purpose behind each code snippet.
    \item \textbf{[S4]} \textbf{Refining and enhancement}: To further enrich the viewing experience, data analyst often refine the recorded content using presentation or video editing software. This process involves removing imperfections, such as minor verbal missteps, and incorporating enhancements like text captions, graphics, or animations to highlight key concepts and make the tutorial more engaging for learners.
    \item \textbf{[S5]} \textbf{Exporting and sharing}: Finally, the meticulously crafted instructional videos are distributed across various online platforms, including learning portals and social media, to reach a wide audience.
\end{itemize}

\subsubsection{Challenges Encountered by Creators} 
\par Following the workflow, we proceed to delineate several challenges (\textbf{C1-C4}) encountered during this process:
\begin{table*}[h]
\centering
\caption{Examples of Narration Classification: 3 Major and 8 Minor Categories}
\label{tab:narration-table}
\begin{tabular}{lll}
\textbf{Contexts} & \multicolumn{1}{c}{\textbf{Examples of Narration}\textit{}} &  \\ 
\cline{1-2}
\textbf{\textit{Informative Text}}  & \textit{-} &  \\
\multicolumn{1}{r}{Background} & \textit{In this video, we're going to explore some real-world economic data using Python and pandas.} &  \\
\multicolumn{1}{r}{Code Interpretation} & \textit{Also, we use Seaborn for data visualization, so let me "import Seaborn as SNS".} &  \\
 & \begin{tabular}[c]{@{}l@{}}\textit{By further narrowing our dataset to "Region == 'Italy'" we ensure that our analysis is }\\\textit{ geographically focused.}\end{tabular} &  \\
\multicolumn{1}{r}{Result Description} & \textit{The result is a pandas dataframe that shows us the series ids.} &  \\
 & \textit{From this bar plot we can see that electronic accessories are purchased in the highest quantity.} &  \\
\multicolumn{1}{r}{Insight} & \textit{There is no relationship between cost of goods sold and ratings.} &  \\
\multicolumn{1}{r}{Conclusion} & \begin{tabular}[c]{@{}l@{}}\textit{Performing EDA, such as univariate, bivariate, and multivariate analysis, allows us to understand }\\\textit{ the underlying patterns and relationships within the dataset.}\end{tabular} &  \\ 
\cline{1-2}
\textbf{\textit{Structural Text}}  & \textit{-} &  \\
\multicolumn{1}{r}{Transition} & \begin{tabular}[c]{@{}l@{}}\textit{We started with importing essential libraries, setting up our environment, and loading the dataset }\\\textit{ to prepare for the exploratory data analysis.}\end{tabular} &  \\
\multicolumn{1}{r}{} & \begin{tabular}[c]{@{}l@{}}\textit{So the next thing we're going to do is try to pull in some data about multiple data series and then }\\\textit{ compare them side by side.}\end{tabular} &  \\
 & \textit{So now our next step in this session we have to get statistics about the new data set.} &  \\
\multicolumn{1}{r}{Direction} & \textit{Pay close attention to the practical application of EDA techniques.} &  \\
 & \textit{Special attention is given to the statistical summaries.} &  \\ 
\cline{1-2}
\textbf{\textit{Interactive Text}}  & \textit{-} &  \\
\multicolumn{1}{r}{Question} & \textit{i'm going to show you how we'll do that here right?} &  \\
 & \textit{Does the cost of goods sold affect customer ratings?} &  \\
\cline{1-2}
\end{tabular}
\end{table*}

\par \textbf{C1. Unclear logical progression}: Creating a video tutorial necessitates a meticulous, step-by-step demonstration. Despite the plannning, the logic is basically based on the code itself, not the planned content of the video. Throughout recording, it's imperative to adhere to a coherent logical sequence, ideally mirroring the logic of the underlying code. However, this aspect is frequently overlooked, with explanations often tied solely to the contents of the notebook. As noted by \textbf{P4}, ``\textit{Sometimes, I catch myself just jumping into the code without giving enough context, making it tough for viewers to follow smoothly.''} To address this, creators could consider briefly pausing or interjecting comments within the code to denote the completion of each phase. As emphasized by \textbf{P3}, ``\textit{Keeping a clear logical flow is key. You've got to lead viewers step by step, making sure it all makes sense with how the code runs.''} Without clear articulation of the logic, viewers may encounter difficulty pinpointing the corresponding part of the video for their code-related issues.

\par \textbf{C2. Lack of visual emphasis in explanations}: Recorded videos often fail to highlight explanations for complex code segments, despite demonstrating the raw process. While verbal emphasis may be given during recording, the scene is typically depicted as a static image, lacking visual aids. As noted by \textbf{P2}, ``\textit{While I was recording, I noticed that even though I stressed things verbally, the absence of visuals might've made it hard for viewers to get it. I might toss in some visuals and animations when I share the clips to help with that.''} Without the reinforcement of visual elements, viewers may struggle to grasp crucial concepts, hindering their understanding and progress. As emphasized by \textbf{P3}, ``\textit{Adding more visuals would've really made the explanations a lot clearer.''}

\par \textbf{C3. Trial-and-error}: Making instructional videos involves a series of trial-and-error processes in recording and editing to achieve clarity and coherence. However, these procedures can be time-consuming, particularly when errors such as slips of the tongue or unexpected interruptions occur during screen recording. As pointed out by \textbf{P1}, "\textit{Editing out errors and unnecessary content really ate up a lot of my time. I mean, there were moments where I stumbled over my words, and unexpected interruptions threw me off track during recording.''} Moreover, recording is essentially irreversible. As noted by \textbf{P4}, ``\textit{You know, when I'm almost done recording and then find an issue with the code I just did, it's such a pain realizing I might have to redo everything from the start.''} Consequently, significant time and effort must be dedicated during the video editing phase to identify and rectify these errors from the outset, ensuring the creation of high-quality content.

\section{Content Analysis}
\label{sec:Narrations}
\par To comprehend the characteristics and design patterns within Jupyter notebook tutorial videos, we embarked on a content analysis. Our first step involved gathering a collection of high-quality digital resources from various online platforms, including \textit{Youtube}~\cite{snickars2009youtube}, \textit{MOOCs}~\cite{saadatdoost2016understanding}, and \textit{Khan Academy}~\cite{thompson2011khan}. We categorized $38$ videos based on their code-cell structure into $372$ sections, each containing the textual narrative and corresponding visuals. Our analysis of Jupyter tutorial videos concentrated on three key areas: (1) categorizing narration, (2) summarizing creators' behaviors, and (3) exploring the correlations between narration and behaviors.

\subsection{Categorizing Narration}

\par The analyzed videos demonstrate similar narrative styles, which are categorized into three major and eight minor categories, with examples provided in \autoref{tab:narration-table}. First, \textbf{\textit{Informative Texts}} offer crucial information to help viewers understand the Jupyter tutorial content.
\begin{itemize}
\item \textbf{Background}: This provides contextual information to help the audience understand the context behind the code cell. For example, ``\textit{we're going to explore real-world economic data using Python and Pandas.''}
\item \textbf{Code Interpretation}: This explains the purpose or function of the code, making it more accessible to viewers. For instance, ``\textit{We're using Seaborn for data visualization, so let's 'import Seaborn as SNS'.''}  
\item \textbf{Result Description}: This clarifies the output of a code cell, explaining what the results indicate. For example, ``\textit{From this bar plot, we can see that electronic accessories are purchased in the highest quantity.''}
\item \textbf{Insight}: This presents the video creator's viewpoints or analyses regarding the results. For instance, ``\textit{There is no relationship between the cost of goods sold and ratings.''} 
\item \textbf{Conclusion}: This summarizes key takeaways and often appears at the end of a significant analysis phase. For example, ``\textit{Performing EDA, such as univariate, bivariate, and multivariate analysis, allows us to understand the underlying patterns and relationships within the dataset.''}
\end{itemize}

\par Second, \textbf{\textit{Structural Texts}} are utilized for organizational purposes, enhancing the video's coherence.
\begin{itemize}
\item \textbf{Transition}: This provides information to help the audience shift focus from one topic to another, e.g., ``\textit{So the next thing we’re going to do is try to pull in some data about multiple data series and then compare them side by side.''}
\item \textbf{Direction}: This directs the audience's attention to specific elements within a cell or code snippets, such as ``\textit{Special attention is given to the statistical summaries.''} 
\end{itemize}

\par Finally, \textbf{\textit{Interactive Texts}} seek to actively engage the audience with the content.
\begin{itemize}
\item \textbf{Question} engages the audience by prompting them to consider upcoming content or outcomes, as in, ``\textit{I’m going to show you how we'll do that here, right.''}  
\end{itemize}

\begin{figure}[h]
    \centering \includegraphics[width=\linewidth]{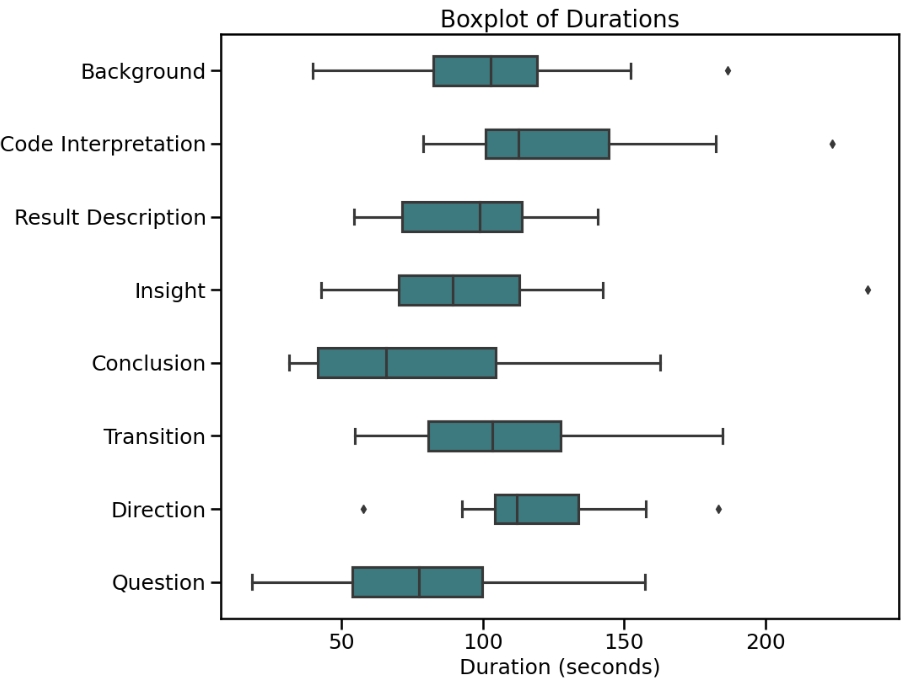}
    \caption{Boxplot of durations for different video segments. Mean durations are as follows: Background at 104.7s, Code Interpretation at 126.7s (the longest), Result Description at 94.9s, Insight at 99.5s, Conclusion at 73.5s (the shortest), Transition at 105.6s, Direction at 118.3s, and Question at 77.4s.}
    \label{fig:video_collection}
\end{figure}

\par During our analysis of the video data, illustrated in \autoref{fig:video_collection}, we observed varying durations for different segments. Notably, Code Interpretation segments stood out as the longest, averaging 126.7 seconds, closely followed by Direction segments, which averaged 118.2 seconds.

\subsection{Summary of Behaviors}
\par 
The creators exhibited a wide range of complex behaviors in the videos. To analyze these behaviors systematically, we conducted a comprehensive summary and examination of each frame within our collected videos. To achieve this, we designed a structured triplet group [sender, receiver, and behavior] to analyze these behaviors. For instance, ``\textit{Bob selected some code snippets in a code unit}'' or ``\textit{Bob elaborated on a code snippet in detail}'' are typical examples of this framework. In this context, the sender refers to the creator, while the receiver encompasses both the notebook and the audience. Through this detailed analysis, we identified a range of behaviors relevant to knowledge sharing and data presentation, omitting actions not directly related to these objectives, such as browsing the internet or taking short breaks. Our analysis primarily centered on the receiver, focusing on both audience engagement and the interpretation of the code presented. Below are the final five behaviors identified:

\begin{itemize}
\item \textbf{Speak}: The instructor verbalizes the narration throughout the data analysis process.
\item \textbf{Live Coding}: The instructor writes, edits, and executes code in real time within a Jupyter Notebook.
\item \textbf{Select}: The instructor highlights key parts of specific pieces of code.
\item \textbf{Annotation}: The instructor adds comments and annotations within the code.
\item \textbf{Pause}: The instructor pauses to prompt the learner to consider what might happen before revealing the outcomes.
\item \textbf{Scroll}: The instructor navigates through code cells one by one.
\end{itemize}

\subsection{Correlation Between Narrations and Behaviors}
\label{sec:patterns}
\par Our interview-based research highlighted the necessity for creators to align their spoken narrations with on-screen actions, especially during the validation of visual components. This synchronization creates a reliable and authentic foundation for showcasing the objectives behind the notebooks. We noticed that while instructors generally maintain a steady flow of narration throughout the process, there are instances during live coding where the continuity of narration may diminish. Through this comprehensive analysis, we identified three key patterns (\textbf{P1-P3}) in the interplay between narration and behavior, which could significantly improve our video creation process:

\par \textbf{P1: Strategic Emphasis through Annotations, Selections, Directions, and Explanations}: In our video tutorials, we strategically utilize \textbf{annotations} and \textbf{selections}, combined with \textbf{directional guidance} and thorough \textbf{explanations} of code segments, to achieve effective emphasis. By using visual and verbal cues to highlight key concepts or critical parts of the code, instructors ensure that important information stands out, making it more memorable for the audience. This emphasis not only captures attention but also helps anchor the audience's focus on crucial points. By providing focused explanations alongside selective highlighting, we effectively guide learners' attention to essential code components, simplifying complexities and clarifying their functionalities.

\par \textbf{P2: Enhancing Cohesion through Transitions and Scrolling}:  Incorporating \textbf{transitions}, such as introducing new coding examples or moving to upcoming topics, ensures a seamless and uninterrupted flow of information. Instructors achieve this by smoothly \textbf{scrolling} from one code cell to the next, maintaining a cohesive and engaging narrative that prevents the tutorial from feeling disjointed or fragmented.

\par \textbf{P3: Encouraging Engagement through Question Prompts and Pauses}: Integrating \textbf{questions} into the narration and providing brief \textbf{pauses} encourages active participation from the audience. This interactive approach stimulates critical thinking and enhances the learning experience by making it more dynamic and engaging.

\section{Design Considerations}
\label{sec:Considerations}

\par Through our formative study and content analysis, we uncovered significant patterns and challenges in creating tutorial videos for conveying analytical processes in computational notebooks. Throughout our discussions, it became clear that relying solely on screen recording formats might not adequately communicate these processes. Our collective deliberations led us to a common objective: to improve the clarity and understanding of the computational notebook's analysis process by tackling the challenges and implementing the identified patterns. This approach entails transitioning from static screen recordings to more interactive and engaging methods, thereby offering a comprehensive insight into the analytical workflows. These insights have led to the following crucial design considerations:

\par \textbf{DC1: Identifying the overview of logical progression.} We aim to meticulously design and construct the script for our ultimate presentation. Resolving \textbf{C1} demands creating a thorough plan that covers all aspects of the presentation content. This plan should include identifying the logical flow of code analysis, highlighting key discussion points, and determining areas of emphasis. The tool should support users in gaining a comprehensive overview of the logical progression.

\par \textbf{DC2: Preserving patterns and enhancing Visualization.} The tool should include features designed to align with the key patterns identified in \autoref{sec:patterns} to foster a deeper understanding and engagement. An example of this is enabling annotations within code cells, enabling users to directly link code segments to explanations, thereby enriching the analytical narrative. Furthermore, the tool should incorporate visual enhancements (\textbf{C2}) like dynamic highlighting or graphical overlays to emphasize these patterns, making the analysis more intuitive and visually appealing.

\par \textbf{DC3: Facilitating rapid preview for expediting creation.} To address \textbf{C3}, the rapid preview feature allows users to promptly evaluate the generated content, ensuring it aligns with their expectations and requirements. This facilitates rapid iterations and adjustments, leading to the creation of high-quality content.

\par \textbf{DC4: Offering flexible editing options for elements.} Building upon \textbf{DC3}, we introduce \textbf{DC4}. When crafting content, refining specific segments is often required after generating the initial version. The system should offer flexible editing capabilities to enable precise fine-tuning, allowing users to seamlessly modify annotations and polish narrations. This ensures users can refine details with ease, ultimately achieving the most optimal presentation in their final output, thus addressing \textbf{C3}.

\begin{figure*}[h]
    \centering
    \includegraphics[width=\linewidth]{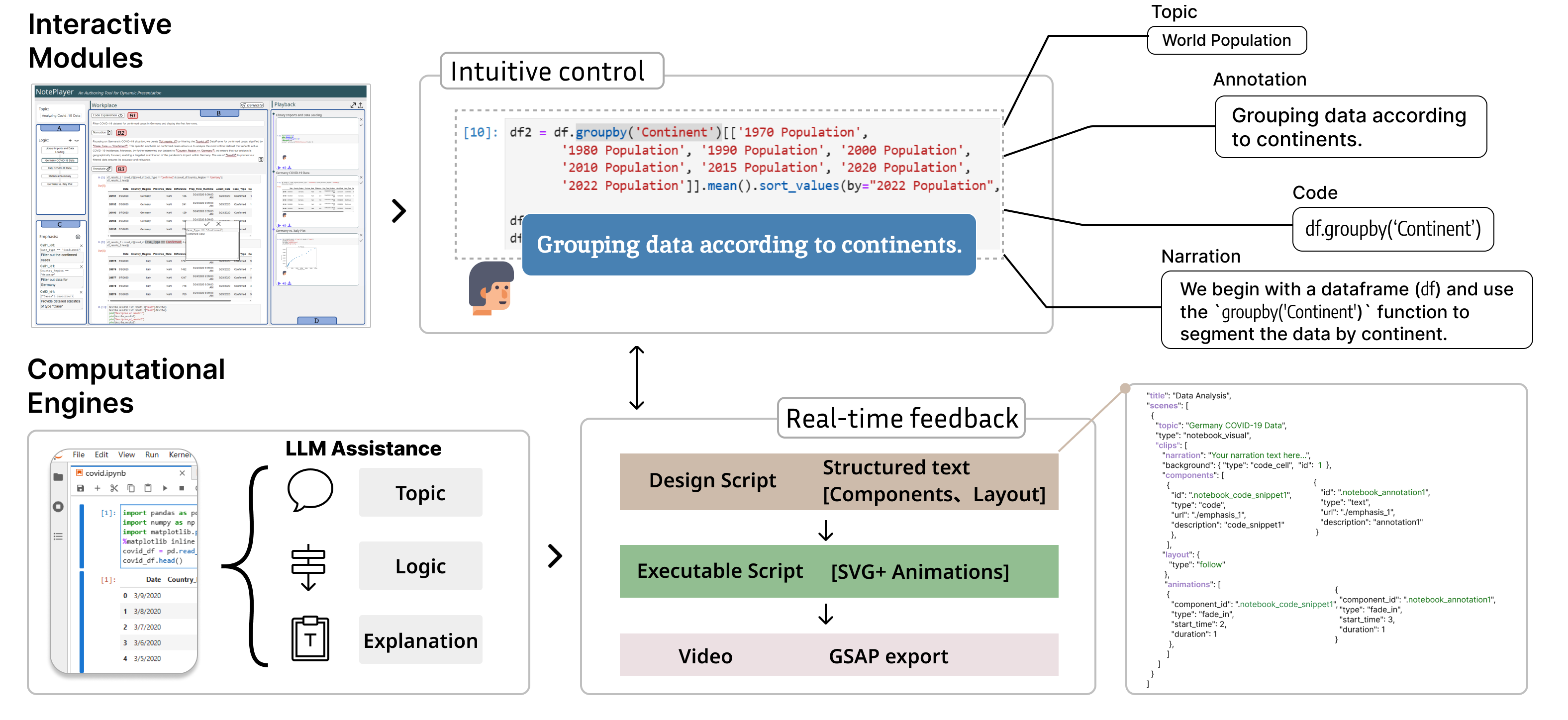}
    \caption{The overview of \textit{NotePlayer} presents two primary components: interactive modules and computational engines. The interactive modules provide intuitive controls and real-time feedback, while the computational engines, serving as the backbone of \textit{NotePlayer}, process data and generate outputs for the interactive modules with assistance from language models.}
    \label{fig:pipeline}
\end{figure*}

\section{NotePlayer}
\par In alignment with our defined design considerations, we develop \textit{NotePlayer}. 
The target of \textit{NotePlayer} is to enhance the seamless and dynamic presentation of the entire analysis process within computational notebooks, facilitating effective communication and understanding between analysts and the notebook environment.

\subsection{Overview}
\label{sec:overview}

\begin{figure*}[h]
    \centering
    \includegraphics[width=\textwidth]{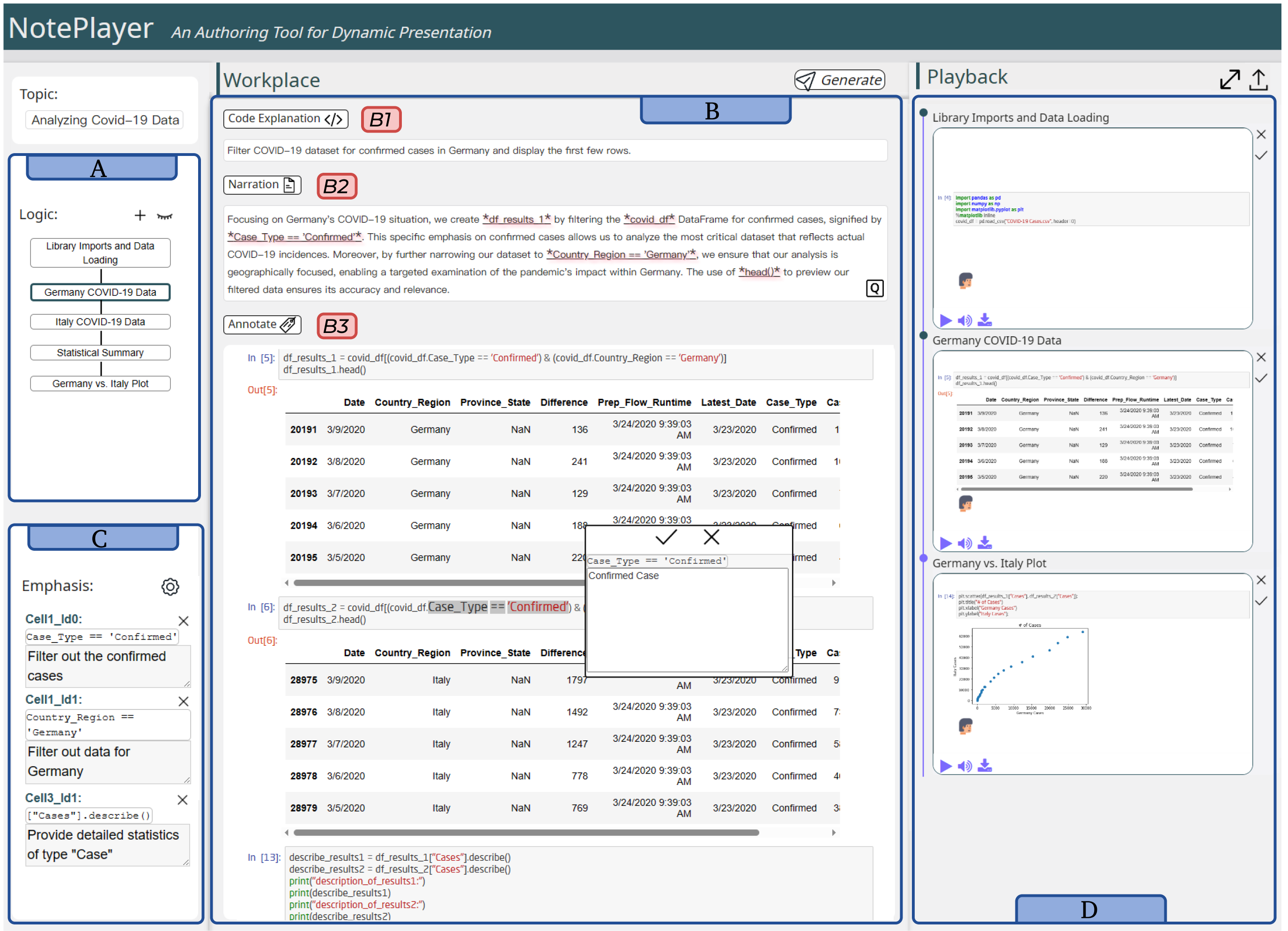}
    \caption{The Interactive Modules of NotePlayer comprise (A) the Logic Flow Representation, (B) the Organization Panel, (C) the Emphasis Record, and (D) Scene Playback. The Organization Panel features crucial elements: (B1) code explanation and (B2) narration section, along with (B3) detailed notebook information, all essential components of the content creation process.}
    \label{fig:interface}
\end{figure*}

\par As shown in \autoref{fig:pipeline}, \textit{NotePlayer} consists of two main components: interactive modules and computational engines, which are intricately interconnected to optimize the user experience. The interactive modules are primarily focused on interface and interaction designs, providing users with intuitive controls and real-time visual feedback. On the other hand, the computational engines serve as the backbone of \textit{NotePlayer}, supporting the functionalities of the interactive modules by processing data, executing algorithms, and generating outputs.

\par To be specific, the interactive modules consist of four parts: the Logic Flow Representation (\autoref{fig:interface}(A)), the Organization Panel (\autoref{fig:interface}(B)), the Emphasis Record (\autoref{fig:interface}(C)) and Scene Playback (\autoref{fig:interface}(D)).
The Logic Flow Representation offers an initial visualization of the logical flow of the notebook code, providing users with a foundation that they can refine and improve upon during the editing phase \textbf{(DC1)}. In the Organization Panel, users have access to essential elements such as code explanations (\autoref{fig:interface}(B1)), narration sections (\autoref{fig:interface}(B2)), and detailed notebook information (\autoref{fig:interface}(B3)), which are crucial for crafting dynamic presentations. Within the Organization Panel, users can fine-tune their content by revising, reorganizing, deleting, and grouping information, contributing to a streamlined editing process (\textbf{DC4}). The Emphasis Record logs sections of the notebook where users apply the "Emphasis" pattern, facilitating focused editing and refinement \textbf{(DC2)}. Both views are interactive and synchronized, ensuring that any modifications made in one view are automatically reflected in the others.
\par The Manipulation Workplace, consisting of the Logic Flow Representation, Organization Panel, and Emphasis Record, provides users with a comprehensive toolkit for refining and organizing elements to create dynamic notebooks. Once users have arranged these elements to their satisfaction, they can preview their work in the Scene Playback (\autoref{fig:interface}(D)) by activating the \raisebox{-0.7ex}{\includegraphics[height=3ex]{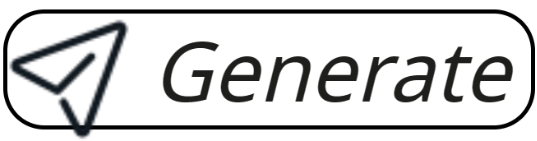}} button located on the right side (\textbf{DC3}).
\par The tool can be seamlessly integrated into Jupyter Lab, allowing for direct incorporation into the notebook environment, which facilitates easy access and sharing.

\subsection{Interactive Modules}
\label{sec:front}
\subsubsection{Logic Flow Representation.}
\label{sec:mani}
This view presents a high-level overview of the notebook's code logic, initially as a straightforward flowchart from top to bottom. It succinctly summarizes the contents of each code cell with a brief description, offering a quick and intuitive understanding of the code's overall flow and functionality \textbf{(DC1)}. Within the Logic Flow Representation, selecting a block offers users two key functionalities: 1) Assess Related Codes: Relevant details of the notebook's code scroll into view within the Organization Panel, allowing scrutiny of the core functions or objectives achieved. 2) Hierarchical Organization: Users are prompted to break down overly complex code blocks into more manageable segments, enhancing clarity and efficiency. Additionally, non-essential blocks can be hidden to maintain focus on relevant code sections.

\par This visualization is designed with two primary considerations: Firstly, it caters to users who, while understanding the basic logic of their codes, benefit from a simplified flow diagram to navigate and locate specific code segments more effectively. Secondly, it empowers users to refine the logical progression, particularly emphasizing the transition from code to video when creating instructional content.

\subsubsection{Organization Panel.}

Within the Organization Panel, users are provided with the convenience of directly accessing the raw codes from their notebooks. This feature facilitates interactive engagement with individual code cells, allowing users to pinpoint specific sections that require attention. Upon selecting a block within the Logic Flow Representation, the corresponding section from the notebook is automatically highlighted, enabling users to interact with and manipulate these codes. Next, we outline the functionalities in alignment with the patterns discussed in \autoref{sec:patterns}: \textit{[F1] Conducting emphasis}, \textit{[F2] Enhancing cohesion}, and \textit{[F3] Fostering interaction} (\textbf{DC2}):

\begin{figure}[ht]
    \centering
    \includegraphics[width=\linewidth]{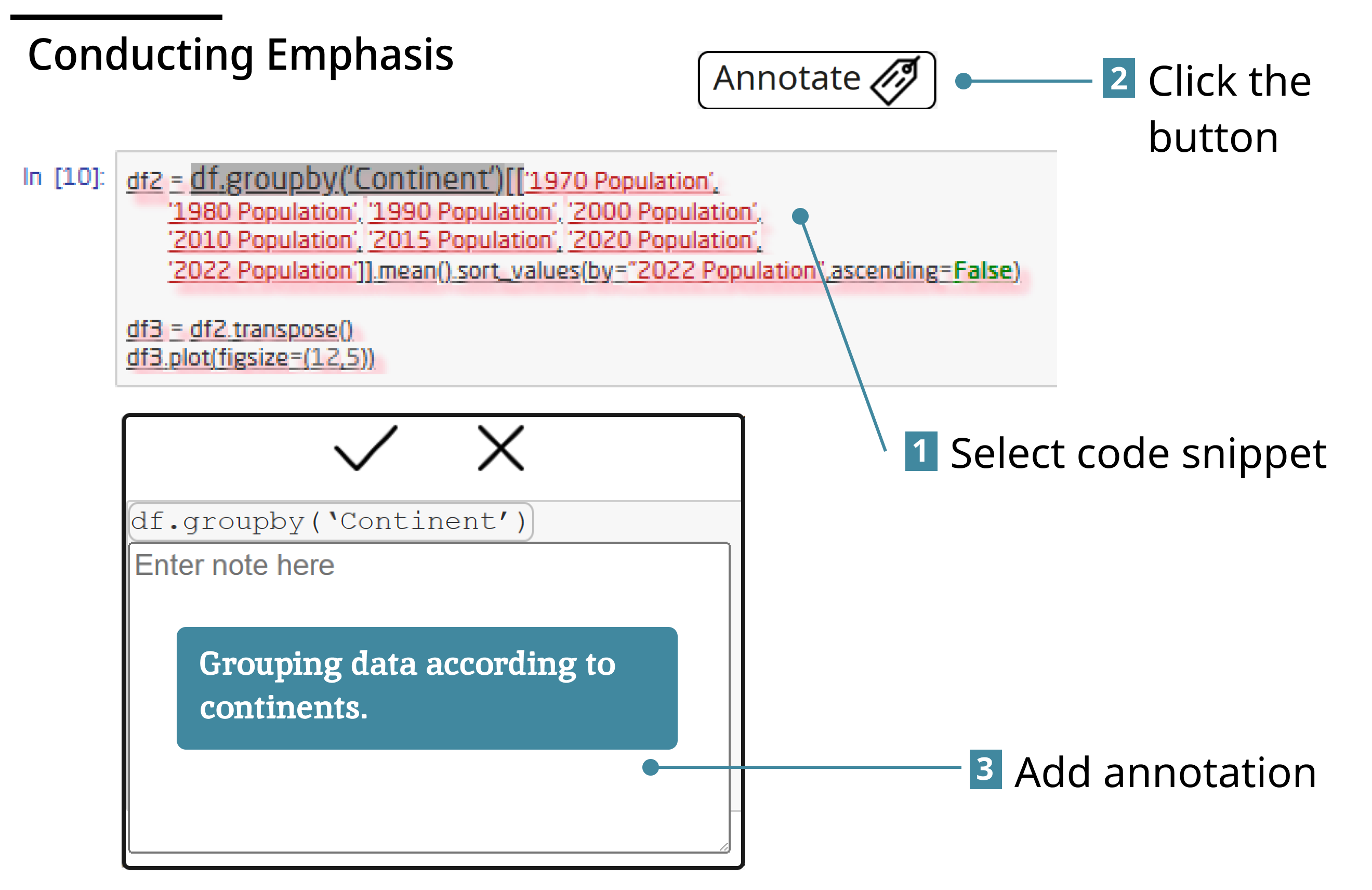}
    \caption{Steps for conducting emphasis: (1) select code snippet, (2) click the button, and (3) add annotation in the pop-up window.}
    \label{fig:emphasis}
\end{figure}
\par \textbf{[F1] Conducting emphasis:} Informed by the formative study and content analysis, we conceptualize these code segments as individual ``scenes'' that users can manipulate, as elaborated in \autoref{sec:ds}. For instance, as shown in \autoref{fig:emphasis}, if users deem the "groupby('Continent')" section within their code as crucial for emphasis during presentations, they can easily highlight this segment using their mouse and confirm their choice by clicking the \raisebox{-0.7ex}{\includegraphics[height=3ex]{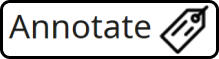}} button. This action increases the font size of the selected code snippet for enhanced visibility and triggers a pop-up window. Within this window, users have the opportunity to elaborate on their reasons or objectives regarding the chosen code snippet. Consequently, an ``emphasis'' visual element is generated, including both the original code and the user's annotations. This process enriches the presentation with valuable context and insights \textbf{(DC2)}.
\par
\begin{figure}[h]
    \centering
    \includegraphics[width=\linewidth]{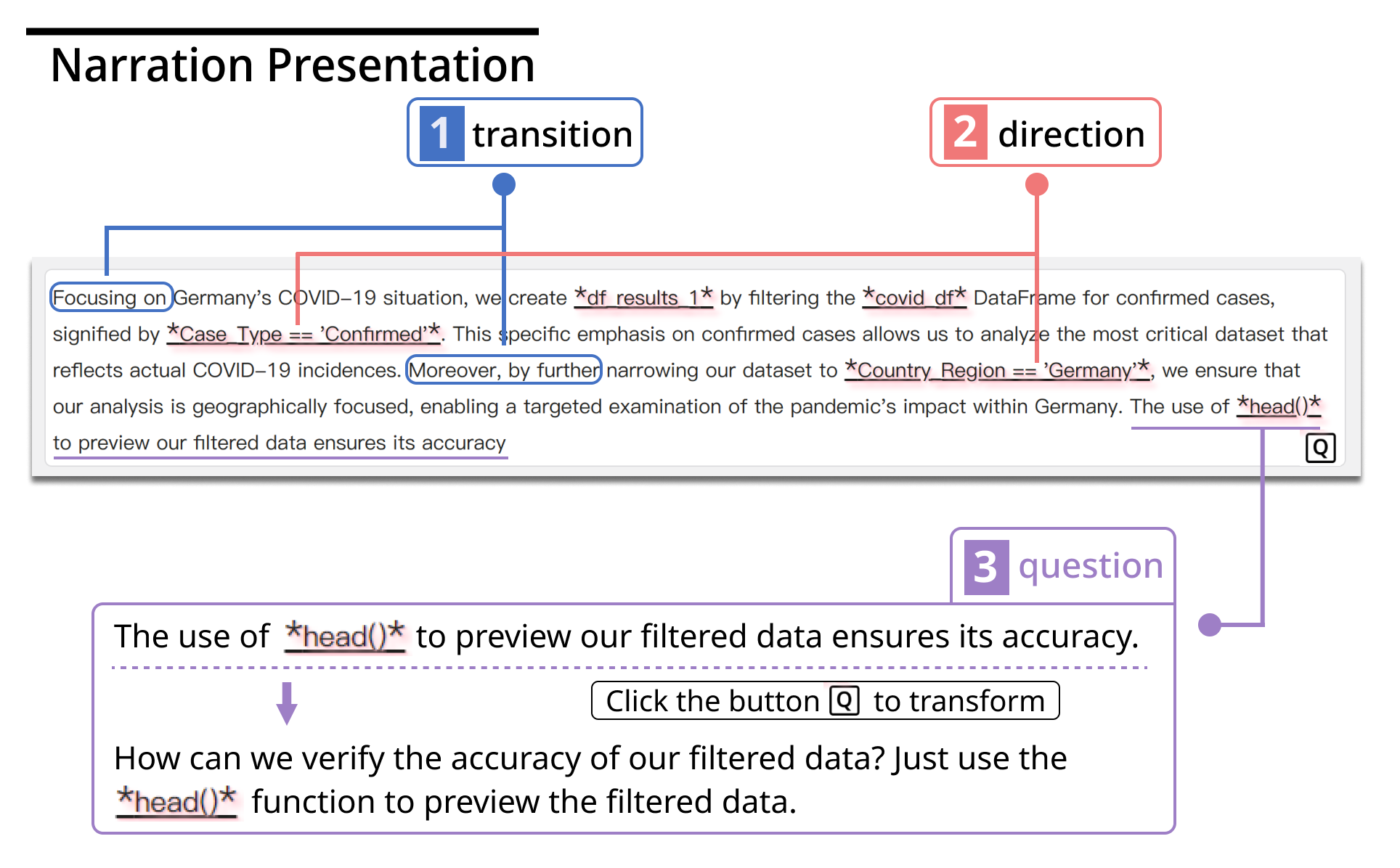}
    \caption{The narration presentation includes three key types: 1) transition, 2) direction, and 3) question. Users can click the \textit{Q} button to transform the sentence from declarative to an interactive question-and-answer format. }
    \label{fig:narration}
\end{figure}

 Once users have thoroughly reviewed a code cell from start to finish, they can choose to generate a corresponding narrative for that segment by clicking the \raisebox{-0.7ex}{\includegraphics[height=3ex]{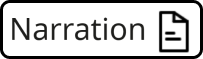}} button. Each sentence in the narrative can be linked with one or two ``emphasis'' visual elements \ouyang{(\autoref{fig:narration}(2))}, providing a detailed explanation or highlighting specific code parts. Users retain the flexibility to modify these narrations \textbf{(DC4)}. By clicking on a sentence, they can access the associated ``emphasis'' visual element, facilitating interactive and in-depth exploration of the code's key points. \revisionyang{This design was naturally derived from our inspection of Pattern 1 in \autoref{sec:patterns}. By using strategic annotations, selections, and directional guidance, we ensure that key concepts and critical parts of the code are effectively highlighted. This method captures attention and anchors the focus on crucial points, simplifying complexities.}

\par \textbf{[F2] Enhancing cohesion:} Ensuring cohesion among code cells is essential for maintaining a seamless analytical process. Efficiently streamlining transitions can be accomplished by strategically placing connecting words like ``focusing on'' or ``by further'' at the beginning and turning points of sections (refer to \autoref{fig:narration}(1)), thus facilitating clear narrative shifts. Moreover, during the narration generation phase, these textual transitions are seamlessly integrated to ensure a coherent flow. Incorporating visual cues, such as ``fade in'' and ``fade out'' animations, not only enhances the narrative experience but also introduces an innovative method for transitioning between code cells. By replacing traditional scrolling with these visual transitions, we can enhance the engagement and intuitiveness of navigating between code cells. \revisionyang{This design aims to maintain cohesion among code cells, aligning with the common practice of organizing content based on cells in many notebook-based tools ~\cite{li2023notable,lin2023inksight,zheng2022telling}, which requires users to effectively structure their notebooks beforehand. Strategically placing connecting words at section beginnings and turning points facilitates clear narrative shifts and ensures coherent flow during narration generation.}

\par \textbf{[F3] Fostering interaction:} Once the narration is generated, it predominantly consists of declarative sentences. However, to enhance user engagement and stimulate analytical thinking, users are provided with the option to select specific sentences and convert them from declarative to an interactive question-and-answer format by clicking the \raisebox{-0.7ex}{\includegraphics[height=3ex]{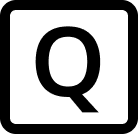}} button. For instance, as illustrated in \autoref{fig:narration}(3), the statement \textit{``The use of *head()* to preview our filtered data ensures its accuracy.''} can be transformed into \textit{``How can we verify the accuracy of our filtered data? Just use the *head()* function to preview the filtered data.''} \revisionyang{Based on the observation of pattern 3 in \autoref{sec:patterns}, we supplement this design to foster interaction. This feature aligns with studies ~\cite{tweissi2016effects,campbell2009questioning} indicating that embedding questions in educational content can enhance learning.}

\par Once users have selected the three specific patterns within a scene, clicking the \raisebox{-0.7ex}{\includegraphics[height=3ex]{figs/generate.png}} button will trigger a preview within the Scene Playback.

\par \revisionyang{We then present \autoref{tab:type-table}, which summarizes the current visual effects and narration strategies in video content, corresponding to the three key purposes: conducting emphasis, enhancing cohesion, and fostering interaction. Design considerations such as animation sets are discussed in \autoref{sec:ds}. Default visual effects are below:} 
\begin{itemize}
\item \revisionyang{\textbf{Fade in}: Gradually introduces elements into the scene, enhancing visibility and focus.}
\item \revisionyang{\textbf{Fade out}: Slowly removes elements from the scene, creating a smooth transition to the next content.}
\item \revisionyang{\textbf{Move to next}: Transitions from the current scene to the next, maintaining viewer engagement through seamless continuity.}
\item \revisionyang{\textbf{Code Snippet Scaling}: Dynamically adjusts the size of code snippets to express emphasis.}
\item \revisionyang{\textbf{Code Snippet Shadow}: Adds a shadow effect to code snippets to increase depth and improve legibility against varied backgrounds.}
\item \revisionyang{\textbf{Annotation Fade in}: Annotations added by the user are smoothly introduced into the video in bullet form, enhancing the explanatory power.}
\item \revisionyang{\textbf{Annotation Fade out}: Gradually fades out annotations, allowing for a clean transition to subsequent content.}
\end{itemize}
\par \revisionyang{Although the current tool features default animations and effects as listed in \autoref{tab:type-table}, future enhancements can include a broader range of animations, a timeline feature for enhanced variability, and non-linear logic flows for smoother transitions across different scenarios.}

\begin{table}[h]
\centering
\caption{\revisionyang{Narration Contexts and Visual Effects for different Purposes in Video Content}}
\label{tab:type-table} 
\resizebox{8.5cm}{!}{
\begin{tabular}{llcc} 
\hline
\textit{\textbf{Purpose}} & \textbf{Description} & \multicolumn{1}{l}{\textbf{Narration Contexts}} & \textbf{Visual Effects} \\ 
\hline
\begin{tabular}[c]{@{}l@{}}Conducting \\ emphasis\end{tabular} & \begin{tabular}[c]{@{}l@{}}Emphasizes key \\codes or content\end{tabular} & \multicolumn{1}{l}{\textit{- Informative Text}} & \multicolumn{1}{l}{} \\
 &  & \multirow{2}{*}{Code Interpretation} & \begin{tabular}[c]{@{}c@{}}Code Snippet Shadow \\\& Scaling\end{tabular} \\
 &  &  & \begin{tabular}[c]{@{}c@{}}Annotation Fade in~\\\& Fade out\end{tabular} \\
 &  & Remain & Fade in \\ 
\hline
\begin{tabular}[c]{@{}l@{}}Enhancing \\ cohesion\end{tabular} & \begin{tabular}[c]{@{}l@{}}Ensures smooth flow \\and continuity\end{tabular} & \multicolumn{1}{l}{\textit{- Structural Text}} & \multicolumn{1}{l}{} \\
 &  & Transition & Move to Next \\
 &  & Direction & Fade in \\ 
\hline
\begin{tabular}[c]{@{}l@{}}Fostering\\interaction\end{tabular} & \begin{tabular}[c]{@{}l@{}}Engage audience \\actively with content\end{tabular} & \multicolumn{1}{l}{\textit{- Interactive Text}} & \multicolumn{1}{l}{} \\
 &  & Question & Fade in \& Fade out \\
\hline
\end{tabular}
}
\end{table}

\subsubsection{Emphasis Record.}
This view is crafted to capture the sections within the notebook that users identify as significant. Once users have identified a segment for video production, they can choose to save these key points \textbf{(DC3)}. This feature streamlines the workflow by providing easy access to important concepts and code snippets for future reference, thus enhancing the efficiency of creating coherent and focused video content \textbf{(DC4)}.

\subsubsection{Scene Playback}
\par Scene Playback presents a sequential preview of user-generated streaming content, aligning each scene with the order depicted in the logic flow representation. To cater to different viewing preferences and maximize space utilization, a ``zoom'' feature is provided. By clicking the \raisebox{-0.7ex}{\includegraphics[height=3ex]{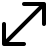}} button, the generated video enlarges for enhanced visibility, and the Organization Panel temporarily hides to expand the playback display area. This functionality enables users to meticulously inspect their content, ensuring it meets their standards with greater precision and clarity. Finally, the `Export' button \raisebox{-0.5ex}{\includegraphics[height=3ex]{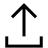}} facilitates the export of all confirmed scene content as an MP4 file.

\subsection{Computational Engines}
\label{sec:back}
\par In this section, we explore the mechanisms through which our computational engines facilitate the production of streaming content. First, we examine the implementation of a ``design script'' that organizes streaming content into coherent scenes, incorporating code, annotations, and visuals. Subsequently, we delve into the role of Large Language Models (LLMs), notably GPT-4, in enhancing narration and logical flow, enabling customization of emphasis to align with user preferences.

\subsubsection{Design Script}
\label{sec:ds}
\par
\begin{figure}[h]
    \centering
    \includegraphics[width=\linewidth]{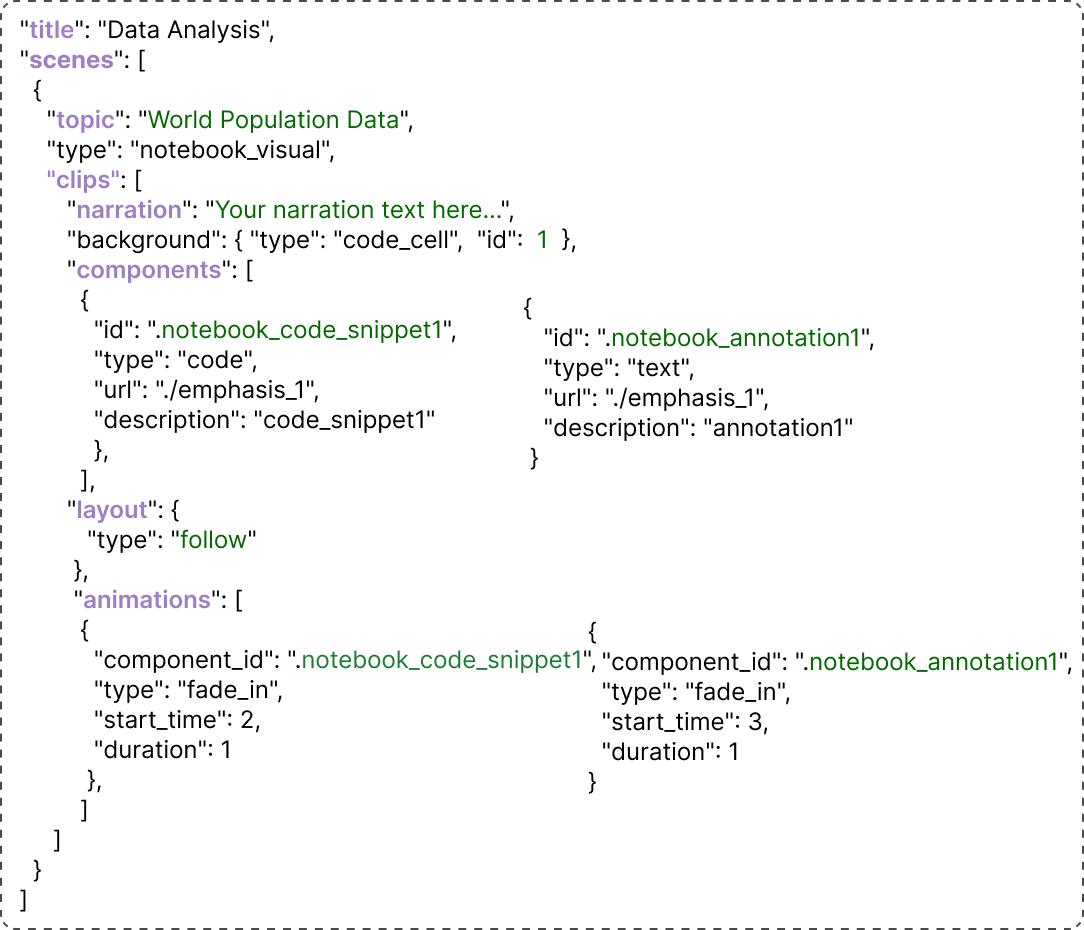}
    \caption{Design script employs a scene-based organizational structure, grouping a narrative segment and its related details into a single block. Each scene comprises a "code snippet," the user's "annotation," and the associated "narrative." This setup is complemented by predetermined sets of animations and default layouts to streamline content creation.}
    \label{fig:design_script}
\end{figure}

 We begin by introducing the concept of a design script, represented as structured text, as illustrated in \autoref{fig:design_script}.

\par \textbf{Scene-based Organization.} The majority of research on storytelling within computational notebooks highlights the use of a section-based architecture for content organization~\cite{wang2023slide4n,wenskovitch2019albireo,harden2023exploring}. This approach segments the narrative into distinct sections, each linking a few concise code cells with explanatory texts and relevant visualizations. Insights gleaned from our formative study interviews and content analysis also indicate the widespread adoption of this structural approach for content streamlining. Drawing on these insights, we adopt a Scene-based organization, where each section of the narrative and its corresponding visuals are grouped within a single block throughout the creating process. This feature facilitates flexible prototyping of the story, visualizations, and animations while ensuring their coherence within each scene~\textbf{(DC4)}.

\par \textbf{Component Details.} The composition of each scene must be clearly defined. Each scene revolves around a particular theme and comprises visual elements accompanied by narration. The core components of this setup are threefold: a ``code snippet'', ``the user's corresponding `annotation''', and ``the corresponding `narrative'''. A single code cell serves as the backdrop for each scene, and a compilation of these foundational elements forms a cohesive scene.

\par \textbf{Layout Sets.} In our formative study, we found that layout considerations are frequently overlooked during recording, leading to suboptimal arrangements that necessitate professional editing for improvement later on. We establish that ``annotation'' in a clip should be positioned near the code cell to ensure coherence and facilitate comprehension.

\par \textbf{Animation Sets.} To address the challenges of developing a comprehensive animation library, we have opted to use a curated set of animations from GSAP (GreenSock Animation Platform)~\cite{gsap}, a widely used animation framework. This approach allows us to concentrate on our primary research goals without incurring the extensive costs of creating a large library. Our focus is on ensuring that the animations serve their intended purpose rather than solely aiming for aesthetic appeal. Drawing from insights gained in our formative study and inspired by existing work on animated storytelling videos~\cite{cheng2022investigating,dataplayer,wonderflow}, we outline the animation format for our content. Each fundamental segment, or ``clip'', is characterized by its start time and duration, which correspond to the application of ``enter'' animations at the outset and ``exit'' animations at the conclusion. Furthermore, within these clips, ``emphasis'' animations are strategically employed to highlight key elements~\textbf{(DC2)}.

\par Once the design script is finalized and confirmed to meet all required specifications, it undergoes processing through an advanced executable program. This program employs Text-to-Speech (TTS) technology to convert the narrations into audio voiceovers, culminating in the creation of the intended streaming content.

\subsubsection{LLM Assistance}
\label{sec:llm}
\par We utilize the advanced natural language understanding capabilities of the LLM (OpenAI's GPT-4 model) to augment the functionality of our tool. \autoref{app:A} provides a detailed list of prompt phrases.
\par \textbf{Logic Flow Generation.} Leveraging GPT's adeptness in code processing and comprehension, this approach guarantees the production of high-quality logic flow representations. Prompt engineering is crafted to direct the LLM in processing notebook cells, utilizing them as input and generating an output organized as a dictionary with the format ``[code-cell index, description, inputs, outputs]''. This output corresponds to each code cell's index, offering a succinct overview of its content, detailing the data or resources it utilizes (inputs), and explaining the outcomes or artifacts it generates (outputs).]

\par \textbf{Narration Generation.} Initially, we experimented with generating narrations directly from raw code cells, treating GPT as an expert in articulating and explaining code~\cite{shanahan2023role}. This experiment demonstrated the model's ability to generate high-quality textual responses. However, recognizing the necessity for customized guidance tailored to diverse user needs, we opted to activate the model only after users highlight their preferred emphasis sections within the Organization Panel, as elaborated in \autoref{sec:mani}. The process takes complete notebooks, enriched with emphasis elements that feature code snippets and user annotations. The output generated includes narrations for each cell, with highlighted ``emphasis'' elements  aimed at effectively conveying the user's intent. Users have the option to adjust and finalize text responses before the generation of streaming content begins.

\begin{figure*}[h]
    \centering
    \includegraphics[width=\textwidth]{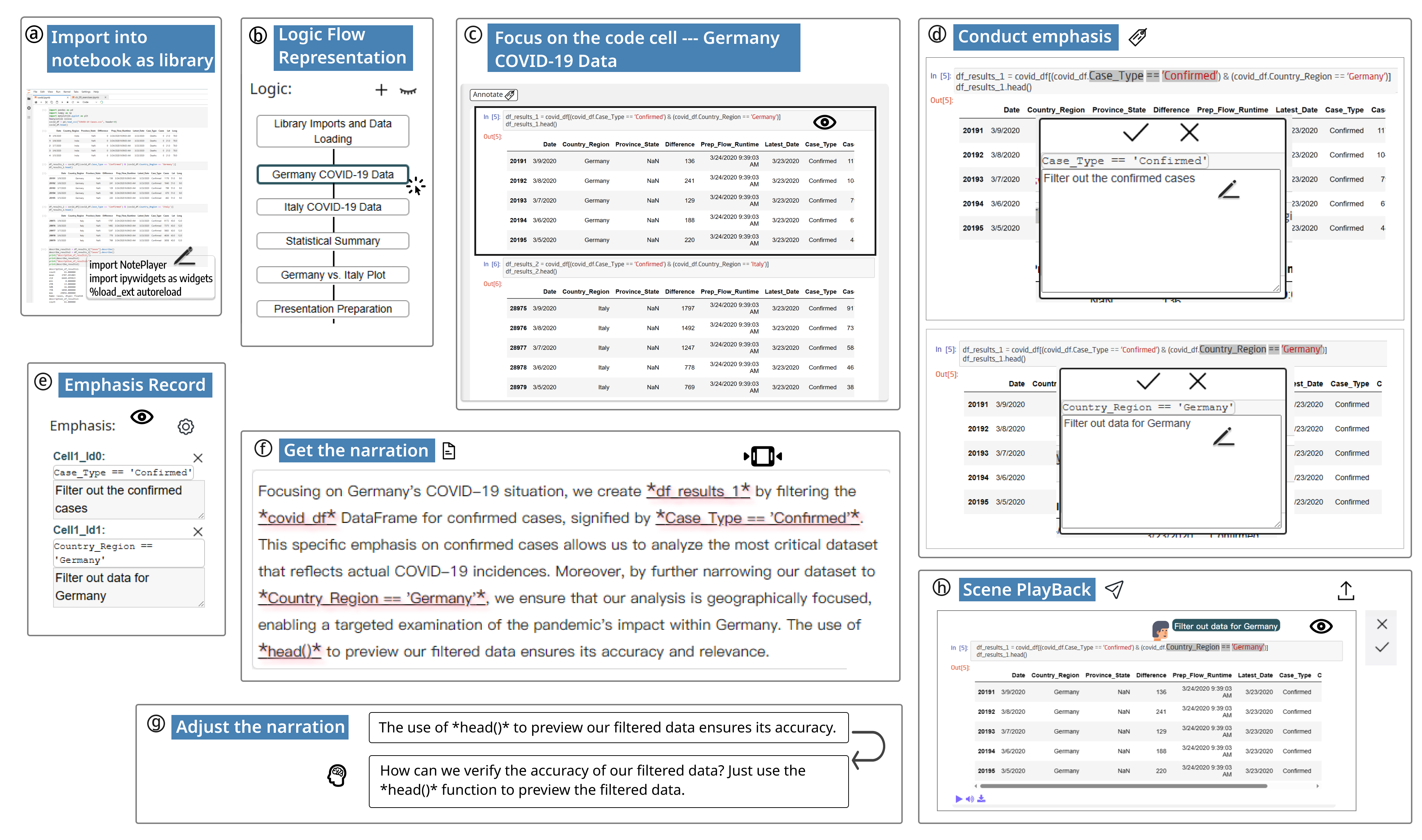}
    \caption{The user's progression in crafting a dynamic presentation using \textit{NotePlayer} follows these steps: (a) integrating the tool into the notebook, then (b) showcasing the Logic Flow Representation. Next, the user (c) directs attention to the ``Germany COVID-19 Data'' code cell and (d) applies emphasis operations. Following this, (e) the Emphasis Record is displayed. Subsequently, (f) the narration linked to the code cell and user annotations is introduced. This is followed by (g) adjusting the narrative style from declarative to interactive question-and-answer format, and finally, (h) demonstrating the Scene Playback.}
    \label{fig:scenario}
\end{figure*}

\section{Evaluation}
\par We evaluated \textit{NotePlayer} through a usage scenario to showcase its expressiveness and a user study to verify its usability.

\subsection{Usage Scenario}
\par We illustrate \textit{NotePlayer}'s capability to dynamically present analytical processes within notebooks through a usage scenario involving Bob, a data analysis professional. While proficient in data analysis, Bob lacks expertise in communication techniques and advanced video editing tools. Typically, he shares his processes by recording his screen and providing verbal explanations.

\par Bob recently concluded a data analysis project on COVID-19 data and aims to share his methodology dynamically. He had finalized the coding in his Jupyter notebooks and outlined a preliminary plan for the presentation's content and structure. Initially, Bob integrated \textit{NotePlayer} into Jupyter Lab as a library (\autoref{fig:scenario}-a). Upon selecting his notebooks, the system generated a logic flow representation based on their contents (\autoref{fig:scenario}-b). This feature offered a succinct summary of each code cell, facilitating a swift understanding of the code's structure and function (\textbf{DC1}). Bob found this representation beneficial for confirming his initial concept of the logic flow. He can inspect each block to access its corresponding code. Satisfied with the descriptions, he chose to make only a few slight adjustments to the names.

\par He concentrated on the ``Germany COVID-19 Data'' code cell (\autoref{fig:scenario}-c), deeming the ``Case\_Type == `Confirmed''' section crucial for his presentation. Bob highlighted this section, enlarged its font size, and annotated it as ``Filter out the confirmed cases''. This action created an ``emphasis'' element (\autoref{fig:scenario}-d), recorded in the \textit{Emphasis Record View} (\autoref{fig:scenario}-e). Similarly, he emphasized the ``Country\_Region == `Germany''' code snippet with an annotation ``Filter out data for Germany'' (\autoref{fig:scenario}-d).

\par After meticulously reviewing the code cell from start to finish, he chose to craft a narrative by engaging the ``Narration'' button (\autoref{fig:scenario}-f). Satisfied with the resulting narration, he believed it adeptly draws attention to the key points he wished to highlight, maintaining coherence and facilitating smooth transitions. Subsequently, Bob pressed the `generate' button to initiate a preview within the \textit{Scene Playback} of \textit{NotePlayer} (\autoref{fig:scenario}-h), showcasing the scene corresponding to the selected code cell. To optimize his view, he utilized the ``zoom'' button, enlarging the video and temporarily concealing the \textit{Organization Panel}. Bob watched the approximately 1-minute-long stream and was content that the presentation effectively conveys his message. He appreciated the synchronization of visual and animation effects with the narration.

\par
\begin{figure*}[h]
    \centering
    \includegraphics[width=\textwidth]{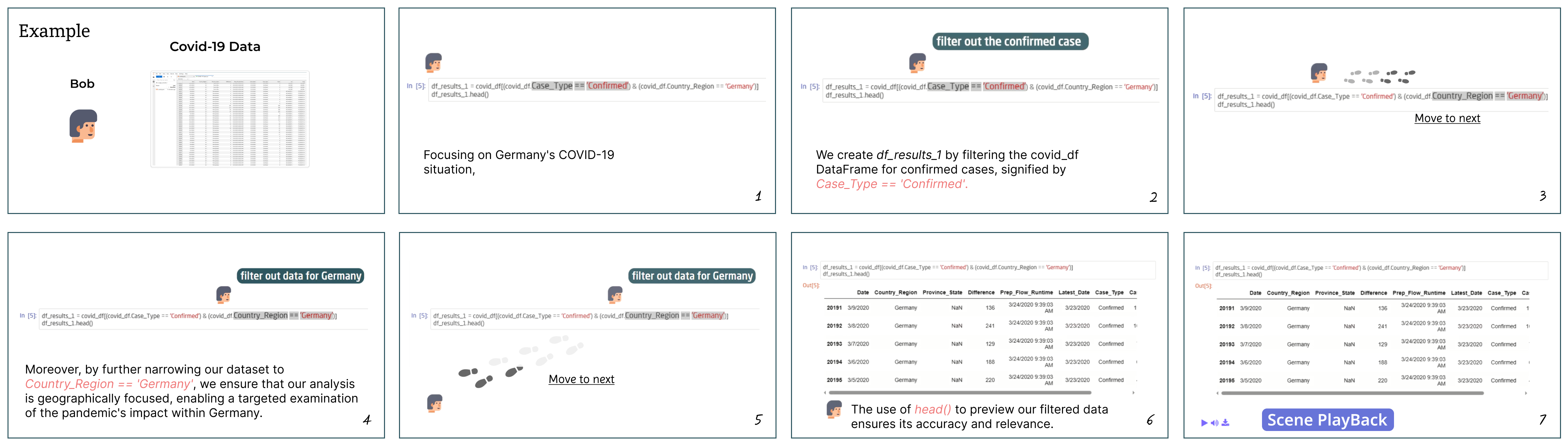}
    \caption{The video example generated for the usage scenario by NotePlayer. The example includes a sequence of annotations and animations. These will be triggered when the audio narration reaches the corresponding segment.}
    \label{fig:example}
\end{figure*}

 Bob, feeling that the narration could be more engaging, decided to improve it. He planed transform the last sentence into an interactive question-and-answer format for increased viewer engagement (\autoref{fig:scenario}-g). After selecting the sentence, he clicked the ``Q'' button, revealing the transformed interactive content. Bob was intrigued by this feature and believed it will enhance viewer interaction with the video. Upon completing the modifications, Bob clicked the ``generation'' button again and promptly observes the updated content in the playback (\autoref{fig:scenario}-h), a process taking just a few seconds. Overall, crafting this small cell scene, along with adjustments, required about 6 minutes to produce a 1-minute stream. Impressed by the efficiency, Bob excitedly exclaimed, ``\textit{Cool. Only 7 minutes have passed. Let’s ship all the notebooks!}''
\revisionyang{Finally, Bob previewed the generated tutorial video shown in \autoref{fig:example}, and was satisfied with its effectiveness.}

\subsection{User Study}
We conducted a user study to verify the effectiveness and usability
of \textit{NotePlayer}.

\subsubsection{Participants}
\par For this study, we recruited 12 participants, comprising five females and seven males, all of whom have experience in sharing Jupyter notebooks, identified as \textbf{P1-P12}. The group includes computer science educators and graduate students with a focus on data analysis. None of the participants had any involvement in the system's design or the preliminary study, and they indicated having minimal to no familiarity with professional video editing tools.

\subsubsection{Materials and Data}
\par We provided the participants with pre-designed notebooks aimed at analyzing a COVID-19 dataset. Alongside the raw notebook, we also supplied them with various pre-extracted key points. This was done to ensure that the user study remained focused on content planning and video creation, thus facilitating clearer communication.

\subsubsection{Procedure}
\par The user study comprises three distinct sessions: (1) a preparatory phase consisting of an introduction and a demonstration example, (2) a creation session allowing participants to engage in the authoring process, and (3) a post-study evaluation to gauge preferences regarding system utility. Each participant completed the entire study within approximately $60-70$ minutes and received a \$15 gift card at the conclusion of the interview session as compensation.

\par \textbf{Preparation.} The user study began with a $15$-minute introduction, elucidating the design objectives of our project. Subsequently, a $20$-minute demonstration was provided, elucidating the utilization of our system, encompassing the setup of visual elements, narrations, and editing interactions within a sample Jupyter notebook. Participants were then encouraged to autonomously explore all functionalities and interactions of the system, with the liberty to pose questions as required. This preparatory phase was designed to furnish participants with the requisite knowledge for proficiently crafting videos using our tool.

\par \textbf{Creation.} After the tutorial, participants were encouraged to utilize our system to craft their own Jupyter notebook videos, leveraging pre-extracted notebook examples for inspiration. They had the flexibility to consult provided slides and request assistance as required. Subsequently, each participant presented and deliberated on their video. This creation phase was allotted a duration of $15$ to $25$ minutes, providing ample time for the task.

\par \textbf{Post-study evaluation.} Once participants concluded their exploration and creation of notebook videos, they were tasked with completing a post-study questionnaire utilizing a $5$-point Likert scale. Here, a rating of $1$ indicated ``strongly disagree'', while $5$ represented ``strongly agree''. The questionnaire's objective was to assess the usefulness, ease of use, and overall satisfaction with our system~\cite{lund2001measuring}. Following this, we conducted semi-structured interviews with each participant to gather qualitative feedback.

\subsubsection{Results and Findings}
\label{sec:findings}
\begin{figure*}[h]
    \centering
    \includegraphics[width=\textwidth]{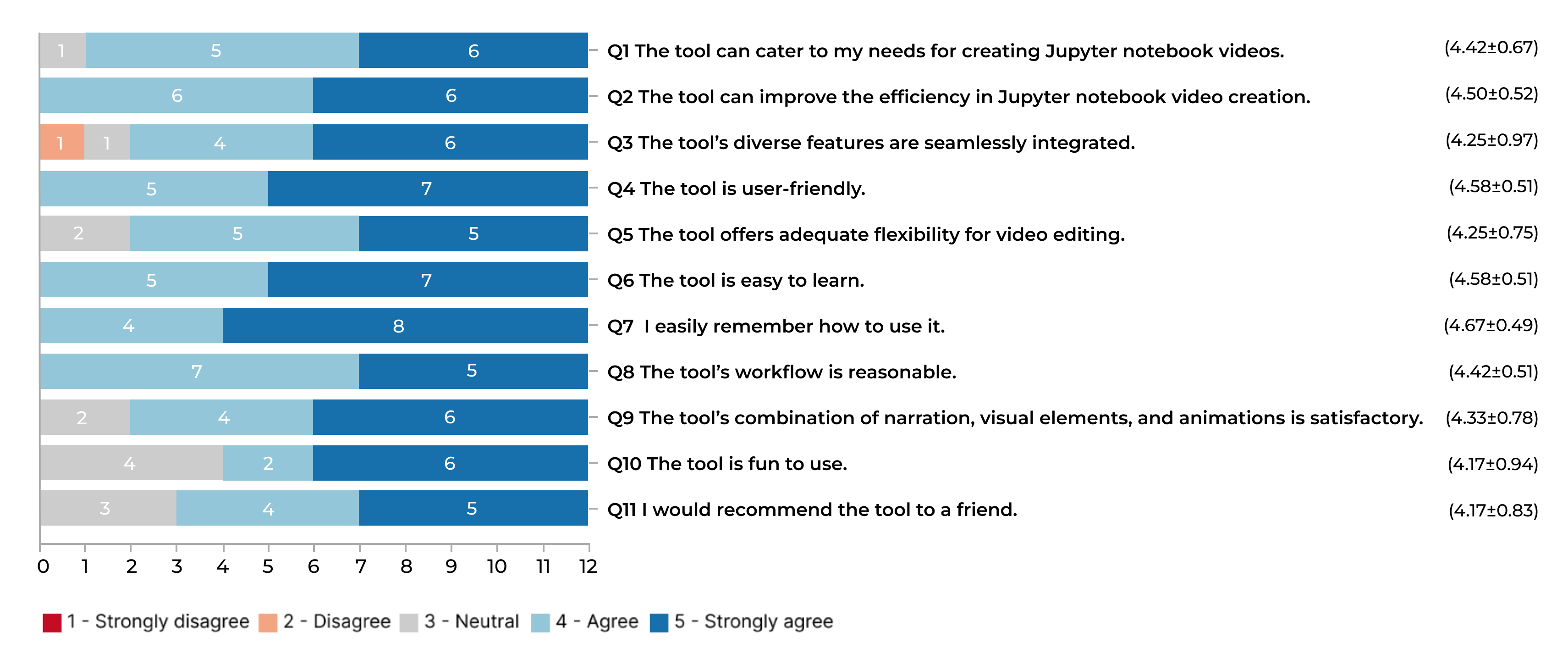}
    \caption{Assessment of our system in terms of \textit{Usefulness} (Q1-Q3), \textit{Ease of Use} (Q4-Q7), \textit{Satisfaction} (Q8-Q11). The middle column shows the detailed questions. The right column displays the average and standard deviations.
}
    \label{fig:user_study}
\end{figure*}

\par We collected participants' subjective evaluations through an 11-question survey and acquired qualitative feedback about our system through semi-structured interviews. \autoref{fig:user_study} illustrates the survey questions along with the average ratings provided by users. In general, participants agreed that \textit{NotePlayer} is an effective tool for creating expressive Jupyter notebook videos, providing clear guidance and being easy to learn. There was a notable preference among participants to utilize \textit{NotePlayer} for streamlining the video prototyping process.

\par \textbf{Usability.} All participants unanimously concurred that our tool significantly streamlines the process of creating programming tutorial videos within notebooks, markedly reducing the necessity for manual efforts. P3 specifically emphasized the tool's usability, stating, ``\textit{It does not require professional video editing capabilities.}'' This semi-automatic approach notably boosts productivity for data analysts. Moreover, participants reported the tool's efficiency in content generation, with P1 remarking, ``\textit{Taking six minutes to generate a one-minute video is very efficient. It can save a lot of time.}'' Additionally, P11 highlighted the tool's efficacy in conveying the user's intention, expressing, ``\textit{I feel the tool is efficient for data analysts, given the intended analytical narrative to be communicated.}''

\par Two participants expressed their preference for the Logic Flow Representation feature. P2 stressed its significance in comprehending the structure of the video, stating, ``\textit{The Logic Flow representation illustrates the distribution of code cells in notebooks to explain an analytical process, providing a clear overview of the video's purpose.}'' Similarly, P4 noted, ``\textit{It's easy to use because it follows the nature 'cell' design of Jupyter notebooks.}''

\par \textbf{Learnability.} The majority of participants found learning and utilizing the integrated features of our system to be intuitive and straightforward. P6, a data analyst with limited experience in video creation and editing, remarked, ``\textit{I can easily learn to share what I'm thinking in my notebook analysis through dynamic means.}'' Additionally, regarding the entire creation process with the tool, many participants suggested that our system ensures a smooth flow in both creating and editing operations. P2 commented, ``\textit{The entire process is smooth and natural.}'' Similarly, P3 mentioned, ``\textit{I can preview the dynamic presentations instantly, with basically no waiting time required.}''

\par \textbf{Expressiveness.} Overall, participants unanimously agreed on the expressiveness of our tool. First, all participants highlighted the value of the preserved pattern. P5 remarked, ``\textit{I like the patterns it contains; it is just what we need for effective communication.}'' Similarly, P6 mentioned, ``\textit{I favor the 'emphasis' pattern the most; it brings the code snippets, user annotations, visual animations, and narrations into a cohesive whole.}'' Second, participants concurred that this dynamic presentation efficiently conveys analytical processes within notebooks. P1 expressed, ``\textit{I'm really into dynamic presentations. They've got the advantages of both screen recordings and slides, but reduce the manual efforts.}'' Adding to this sentiment, P3 stated, ``\textit{I think screen recording is really troublesome, and I have to talk through it all and then edit it. The narrative summary and generation of the tool is expressive to me.}'' Third, multiple participants emphasized the importance of quickly assessing whether their intention made sense or not. \textit{NotePlayer} offered a unique way to streamline content creation through the use of preset settings. As noted by P8, ``\textit{Sometimes I am not sure whether to use position, color, or animation to convey the dynamic process... it can be hard to think of, I just want to emphasize some code snippet.}'' For data analysts, effectively communicating information takes precedence over special effects.

\par \textbf{Flexibility.} Over half of the participants commended the flexibility of our tool in creating Jupyter videos. P4 highlighted the freedom the tool provides for adding and editing user annotations, contrasting it with previous tools that required constant switching between editing interfaces and notebooks. Similarly, P10 was impressed by the scene organization and the design script, which facilitated the reuse of streaming, thereby enhancing the tool's utility. Additionally, P5 remarked, ``\textit{With LLM's support, I can easily refine the narrations, which are flexibly linked with specific code snippets and annotations, all defined together as a scene.}''

\par \textbf{Suggestions.} During our study, participants highlighted a few areas where \textit{NotePlayer} could potentially enhance to make the video creation process more adaptable and user-friendly. First, the current linear presentation adheres to the sequence of scenes, reflecting the inherently linear nature of the video's structure. However, some participants expressed a desire to adjust the logic flow representation from linear to a tree structure. P8 noted, ``\textit{It's fine to have the logic go step by step, but some code does similar things and the tree structure really helps me understand the flow better.}'' Second, several participants voiced a desire to incorporate a timeline feature for more granular control over the streaming content. As pointed out by P12, ``\textit{A timeline feature, similar to that in Adobe Premiere Pro, could be more beneficial, allowing for a quick overview of the total duration of each element within one scene.}'' Last, some participants mentioned their interest in experimenting with a wider range of animations. As stated by P11, ``\textit{Trying different animations in dynamic presentations could be pretty fun.}''

\section{Discussion and Limitation}
\par In this section, we reflect on the design, implementation, and evaluation of
\textit{NotePlayer}, and explore potential research avenues and future directions.

\subsection{Extend to Multifaceted Authoring Scenarios}
\par While the user study revealed positive feedback regarding the usability of \textit{NotePlayer}, its design may not fully accommodate the diverse authoring scenarios. Originally conceived as a plugin, technical constraints and the need for enhanced convenience, such as simplified positioning and pop-up effects, led us to develop an interface integrated with Jupyter Lab. Our plan includes reintegrating the generated dynamic presentations back into the raw code cells embedded in the output cells, ensuring a seamless experience within the notebook environment. This approach allows users to view streaming of smaller scenes if they have questions about specific code cells. Second, our assumption regarding the workflow—that users have pre-organized and optimized their notebook before creating a dynamic presentation—may be overly simplistic. Previous research suggests that selecting and organizing notebook cells is a complex process, indicating the need for our tool to include functionalities that better support these tasks. Third, our tool currently lacks complete user control over all design aspects, such as animation styles and annotation colors. While only a few participants raised this issue in our user study, the majority found the provided options sufficient for emphasizing key points and facilitating communication. Moving forward, we aim to introduce more personalized settings and multimodel interactions to enhance the videos' appeal and stylistic diversity, catering to a wider range of users~\cite{vistalk,NLISurvey}.

\par \revisionyang{Furthermore, extending tutorials to other domains is an exciting prospect, as demonstrated by Chi's work~\cite{chi2013democut}, which highlights the need to customize tutorial tools for specific domains. \textit{NotePlayer} enhances programming tutorials by merging code with video, automating narration, and integrating interactivity. Similarly, \textit{DemoCut}~\cite{chi2013democut} automates editing for physical tutorials using user annotations and visual effects, significantly reducing manual effort. While both tools streamline tutorial creation, \textit{NotePlayer} excels in digital interactivity, whereas \textit{DemoCut} enhances continuous physical actions. Developing these tutorials requires a clear structure and engaging content, though their technical requirements and solutions differ. Nevertheless, methodologies from different domains can inform and enhance each other, offering avenues to explore~\cite{emerson2024anther}. For instance, the emphasis on tutorial videos across various fields can differ, enabling us to investigate and experiment on how diverse designs impact learners' views of usage patterns and result in improved learning results.}
\revisionyang{Additionally, in data analysis scenarios that heavily rely on Excel, \textit{NotePlayer} could dynamically present and narrate Excel-based processes, including cases where complex formulas or codes are used for analysis within Excel~\cite{bree2016using}. Moreover, certain outputs generated by our tool, organized through a cell-by-cell division process~\cite{li2023notable,zheng2022telling,wang2024outlinespark}, can seamlessly integrate into PowerPoint, Keynote, and Google Slides presentations. Furthermore, key interpreters, animations, and narrations are envisioned as potential future plugins, with the prospect of embedding them into IDEs like Visual Studio Code~\cite{bin2022mastering}.}
\subsection{Serve as a Complement Format}
\par \revisionyang{Tutorial videos provide an additional layer of support, effectively demystifying complex topics, increasing engagement, and offering flexible learning options. However, it's crucial to acknowledge that video-based learning may not suit all learners, as noted by Wells's work \cite{wells2012using}. Our intention is not to replace traditional approaches but to supplement them.} We also recognize the substantial value and widespread use of traditional screen recording in educational settings, training programs, and online learning~\cite{abdous2010learner,luongo2015missing}. Traditional screen recording significantly enhances instructor presence and helps bridge the gap between educators and learners in tutorial videos~\cite{Kelly2018Using}. It allows for the detailed capture of desktop actions, facilitating hands-on teaching activities like live coding. This aids in the clear demonstration of new technologies to student~\cite{Daniels2009Technically,Bernal-C'ardenas2020Translating}.

\par Our focus lies in providing an intuitive and efficient solution tailored for data analysts, specifically designed for scenarios involving the exchange of analytical processes. Conventional screen recording formats may not always cater effectively to this audience. Significantly, our approach draws inspiration from conventional screen recording practices. We assimilate and integrate discerned patterns (refer to \autoref{sec:patterns}) from these methods to enhance our content delivery.  \revisionyang{Our narrative styles analyzed in \autoref{sec:Narrations} also align with the elements that support video editing as outlined in prior studies~\cite{yang2023beyond,chi2013democut}. In fact, this taxonomy can enhance users' viewing experiences by helping them quickly locate and skip irrelevant information based on the category and type, as indicated by these studies. We can further enhance our system by adding features like Opening/Closing or self-promotion type narrative information to accommodate users with diverse information needs and navigation preferences~\cite{yang2022improving}. By doing this, we expect video creators to have more control, allowing them to make informed decisions and better serve as a complementary format~\cite{jin2017elasticplay}.} Furthermore, our user study revealed that participants readily comprehend the workflow and identified patterns within our tool, finding them intuitive and familiar (refer to \autoref{sec:findings}). Looking ahead, our objective is to incorporate more advantages of traditional screen recording into our tool, aiming to enhance its expressiveness. This entails enriching videos with personalized and captivating elements~\cite{sablic2021video, morton2016blended}, and embracing innovative presentation styles such as Learning Glass~\cite{choe2019student}.
\par

\subsection{Incorporate LLM Assistance}
\par We employed LLMs (refer to \autoref{sec:llm}) to assist in generating logic flow and creating narrations. On one hand, the logic flow content generated consistently proved accurate, delivering significant value to participants in our user study. Conversely, our approach to generating narrations, which aimed to align with user intentions, also garnered positive feedback during the study. Despite their positive performance, current LLMs are constrained by inherent limitations. These include inconsistency in outputs across different iterations and issues with generating inaccurate or fabricated information (hallucinations)~\cite{ji2023towards,huang2023survey}. It may be prudent to implement a narration revision record to monitor changes. Moreover, although users have the option to refine narrations and transition from a declarative to an interactive question-and-answer format, two participants specifically expressed a desire to present notebook narrations in their own voice or style. This preference underscores the necessity for more personalized narration options, such as fine-tuning the model using materials provided by users themselves, tailored to individual preferences~\cite{hu2023llm}.

\subsection{Limitations}
\par Our work has several limitations. First, our user study involved a limited number of participants, underscoring the need for a larger, crowdsourced study to comprehensively evaluate the usability of our tool. Additionally, the absence of a baseline in our study for comparison is noteworthy. While participant feedback indicated satisfactory communication of their analytical processes through streaming, a quantitative comparison could offer deeper insights into these initial findings. Second, in our study, we examined essential design considerations based on our formative study and content analysis. It is important to recognize that these considerations serve as foundational guidelines to assist data analysts in video production and do not constitute an exhaustive list. Although we analyzed $38$ videos and identified three distinct patterns, varying scenarios may necessitate more nuanced strategies to effectively address their specific requirements. For instance, code cells dedicated to displaying results should prioritize highlighting these outcomes, while those involved in data manipulation may require additional emphasis to enhance the explanation and guidance provided for these segments. 
\revisionyang{Moreover, while our tool currently excels in illustrating the analytical process, it lacks capabilities for hands-on teaching scenarios where progressive code display is essential. To address this, we plan to enhance our tool by introducing features that allow for the customization and selection of specific code cells, as well as progressive code display for focused presentations. These improvements will enhance the versatility and effectiveness of code displays across various use cases. Additionally, incorporating recommendation features based on customer feedback can significantly enhance the quality of tutorial videos. We plan to conduct a user study to gather insights on customer satisfaction and the learning experience with our methods~\cite{yang2022softvideo}. By using statistical analysis to identify trends, we aim to provide personalized guidance, which could greatly increase both customer satisfaction and engagement.}

\section{Conclusion and Future Work}
\par This study offers valuable insights into the authoring of video tutorials within Jupyter notebooks, focusing on the analytical process. We introduce \textit{NotePlayer} as a solution to overcome the challenges faced by users in this context. Through an extensive formative study and content analysis, we have identified significant patterns and obstacles in the creation workflows, underscoring the necessity for innovative tools to improve user experience and communication. \textit{NotePlayer} represents a substantial advancement in this domain by seamlessly integrating video segments with notebook cells, thus enabling more engaging and flexible communication of analyses. Our user study further validates the effectiveness of \textit{NotePlayer}, highlighting its benefits while also pinpointing areas for refinement. Moving forward, there are several avenues for future exploration and enhancement. Incorporating user feedback to expand features, such as enhancing editing capabilities and providing personalized options, could augment both utility and usability. Moreover, conducting long-term evaluations in real-world settings could yield deeper insights into its practical applications.

\begin{acks}
We thank all our participants for their time and valuable input. We would also like to thank our reviewers whose insightful comments have led to a great improvement of this paper. This work is supported by grants from the National Natural Science Foundation of China (No. 62372298), the Shanghai Frontiers Science Center of Human-centered Artificial Intelligence (ShangHAI), and the Key Laboratory of Intelligent Perception and Human-Machine Collaboration (ShanghaiTech University), Ministry of Education.
\end{acks}

\bibliographystyle{ACM-Reference-Format}
\bibliography{sample-base}

\newpage

\appendix

\onecolumn

\section{Prompt Design}
\label{app:A}
\par For these two tasks, we adhere to the following prompt engineering principles proposed by OpenAI~\cite{gptpromptengineer}:

\begin{enumerate}
    \item \textbf{Include details in your query to get more relevant answers.} By explicitly listing the key considerations, steps to complete the task, and output requirements in the task description, we assist the model in understanding and executing the task more accurately.
    \item \textbf{Ask the model to adopt a persona.} We explicitly specify the role of the model in the prompt, aiding the model in better understanding the task's context and requirements, thereby providing more accurate and relevant answers. Through persona setting, the model can resonate better with the task's content and demands, enhancing the effectiveness and quality of task execution.
    \item \textbf{Use delimiters to clearly indicate distinct parts of the input.} We employ clear delimiters and headings, such as ``\textit{Key Considerations}'', and ``\textit{Steps to Complete the Task}'', in the prompt. This aids the model in understanding and organizing the input information more clearly, thereby improving task efficiency and accuracy.
    \item \textbf{Specify the steps required to complete a task.} We explicitly list the specific steps required to complete the task in the prompt, such as ``\textit{read and understand the sentences}''. This provides clear operational guidance to the model, aiding in enhancing task execution efficiency and quality.
\end{enumerate}
The detailed prompts designed for interaction with GPT-4 are presented below.
\subsection{Logic Flow Generation}
\begin{lstlisting}
{
    "role": "system", "content": 
     """
    # Task Details
    **Objective**: 
    You are a code analyst. Your task is to analyze the logic flow within a series of notebook cells to construct a comprehensive representation of the computational logic. This involves identifying the purpose and functionality of each cell, the inputs it relies on, and the outputs it generates.

    **Key Considerations**:
    - Understand the specific role of each notebook cell within the overall computational process.
    - Identify the inputs required by each cell and the outputs it produces.
    - Recognize any dependencies between cells, noting how the output from one cell might serve as input to another.

    **Steps to Complete the Task**:
    1. Examine each notebook cell to discern its primary function and role in the notebook's logic flow.
    2. Catalog the inputs each cell uses, which may include data files, variables from previous cells, or user inputs.
    3. Determine the outputs each cell generates, such as data transformations, visualizations, or results.
    4. Construct a logic flow representation, organizing this information into a structured format.

    **Output Requirements**:
    - Provide the logic flow representation in JSON format.
    - Represent each cell as an object within the JSON structure, including the following key-value pairs:
      - 'id': A unique identifier for each cell (starting from 0, incrementing by 1 for each subsequent cell).
      - 'description': A brief explanation of the cell's purpose and functionality.
      - 'inputs': A list of the inputs the cell uses.
      - 'outputs': A list of the outputs the cell produces.

    **Attention**:
    Ensure that the JSON representation accurately reflects the sequence and dependencies of the notebook cells, offering clear insights into the notebook's computational logic.

    ----
    # Output Example
    ```
    [
    {"id": 0, "description": "Load dataset", "inputs": ["data.csv"], "outputs": ["raw_data"]},
    {"id": 1, "description": "Clean data", "inputs": ["raw_data"], "outputs": ["cleaned_data"]},
    ...
    ]
    ```
    """
}
\end{lstlisting}

\subsection{Narration Generation}
\begin{lstlisting}
{
    "role": "system", "content": 
     """
    # Task Details
    **Objective**: 
    As a code_interpreter, your mission is to generate narrations for notebook cells, particularly focusing on user-highlighted code snippets ("emphasis" elements) and their corresponding annotations. A key part of your role is to weave these individual narrations into a coherent overall narrative that reflects the interconnected functionality of the notebook cells.

    **Key Considerations**:
    - Integrate user-highlighted code snippets and annotations into the narrations, providing depth and insight into specific functionalities.
    - Ensure coherence among cell narrations, understanding the function of each cell in relation to others, to maintain a seamless narrative flow across the notebook.
    - Reflect both the technical essence of each cell and the user's perspective, enhancing the narrative with insights into why certain code snippets are emphasized.

    **Inputs for Each Cell**:
    - **Code Snippet**: The specific portion of code highlighted by the user as significant.
    - **Annotation**: The user's explanation for highlighting this snippet, offering additional context or importance.

    **Steps to Complete the Task**:
    1. For each notebook cell, identify the emphasized code snippet(s) and the corresponding user annotations.
    2. Craft a detailed narration for the cell that not only explains its specific functionality but also incorporates the user's emphasis, providing a richer understanding of its role.
    3. In generating narrations, consider the functions of adjacent cells to ensure that your narrative offers a coherent explanation of how each cell contributes to the notebook's overall logic flow.
    4. Assemble the individual narrations into a structured format that reflects the interconnectedness and sequence of the notebook cells, enhancing narrative flow.

    **Output Requirements**:
    - Narrations should be formatted in JSON, with each object representing a cell's narration and including:
      - 'id': The unique identifier for the cell.
      - 'narration': The cell's detailed explanation, integrating the emphasized code snippets and annotations.
      - 'inputs': The emphasis elements and annotations guiding the narration.
    - The narrations must collectively offer a coherent narrative across the notebook, elucidating the interconnected functionality of the cells.

    **Attention**:
    It's crucial that the narrations not only individually explain the functionality of each cell but also collectively contribute to a coherent narrative of the notebook's computational logic. This involves understanding the broader context in which each cell operates and how they interrelate.

    ----
    # Output Example
    ```
    [
      {
        "id": 0,
        "narration": "This cell initiates our data analysis by loading data from 'data.csv', focusing on the 'read_csv' function. Highlighted by the user, 'read_csv' is crucial for efficient data loading, preparing us for subsequent preprocessing steps.",
        "inputs": {
          "code_snippet": "data.read_csv('file.csv')",
          "annotation": "Essential for initial data loading."
        }
      },
      {
        "id": 1,
        "narration": "Following data loading, this cell applies 'dropna()' to clean the dataset, a step emphasized by the user for its importance in ensuring data quality. This cleaning is foundational for the analysis performed in later cells.",
        "inputs": {
          "code_snippet": "data.dropna()",
          "annotation": "Crucial for maintaining data integrity, leading to reliable analysis."
        }
      }
    ]
    ```
    """
}
\end{lstlisting}

\end{document}